%% file: Martin_et_al_JCGS_revision_final_18-10-29.tex
\pdfoutput=1
\documentclass[12pt]{article}%
\usepackage[left=2.1cm, right=2.2cm, top=1.98cm, bottom=2.1cm]{geometry}
\usepackage{amsfonts}
\usepackage{amssymb}
\usepackage{amsmath}
\usepackage{lscape}
\usepackage{color}
\usepackage{algorithm}
\usepackage[noend]{algpseudocode}
\usepackage{graphicx}
\usepackage{tikz}%
\providecommand{\U}[1]{\protect\rule{.1in}{.1in}}
\newtheorem{theorem}{Theorem}

\makeatletter
\def\BState{\State\hskip-\ALG@thistlm}
\makeatother
\begin{document}

\title{Auxiliary Likelihood-Based Approximate Bayesian Computation in State Space
Models\thanks{This research has been supported by Australian Research Council
Discovery Grants No. {DP150101728 and DP170100729}. Supplementary material for
this article (including computer code) is available on-line. We thank the
Editor, Associate Editor and three anonymous referees for very detailed and
constructive comments {on earlier drafts of the paper}.}}
\author{Gael M. Martin\thanks{Department of Econometrics and Business Statistics,
Monash University, Australia. Corresponding author; email:
gael.martin@monash.edu.}, Brendan P.M. McCabe\thanks{Management School,
University of Liverpool, U.K.}, David T. Frazier$^{\text{\textdagger}}$,
\and Worapree Maneesoonthorn\thanks{Melbourne Business School, University of
Melbourne, Australia.} and Christian P. Robert\thanks{Universit\'{e}
Paris-Dauphine, Centre de Recherche en \'{E}conomie et Statistique, and
University of Warwick.}}
\maketitle

\begin{abstract}
A computationally simple approach to inference in state space models is
proposed, using approximate Bayesian computation (ABC). ABC avoids evaluation
of an intractable likelihood by matching summary statistics for the observed
data with statistics computed from data simulated from the true process, based
on parameter draws from the prior. Draws that produce a `match' between
observed and simulated summaries are retained, and used to estimate the
inaccessible posterior. With no reduction to {a low-dimensional set of
}{sufficient statistics} being possible in the state space setting, we define
the{\ summaries} as the maximum of an auxiliary likelihood function, and
{thereby exploit }the asymptotic sufficiency of this estimator for the
auxiliary parameter vector. We derive conditions under which {this approach -
including a computationally efficient version based on the auxiliary score -
achieves Bayesian consistency}. To reduce the well-documented inaccuracy of
ABC in multi-parameter settings, we propose {the separate treatment of} each
parameter dimension {using an integrated likelihood technique}. Three
stochastic volatility models for which exact {Bayesian} inference is either
{computationally} challenging, or infeasible, are used for illustration.{ We
demonstrate }that our approach compares favorably against an extensive set of
approximate and exact comparators. An empirical illustration completes the
paper.\medskip


\end{abstract}

\section{Introduction}

\baselineskip21pt

Approximate Bayesian computation (ABC) serves as an effective, and sometimes
unique tool, for conducting inference in models with intractable likelihoods,
with there being no restriction on the area of application. (See Marin
\textit{et al}., 2011, and Sisson and Fan, 2011, for reviews.) The technique
circumvents direct evaluation of the likelihood function by selecting
parameter draws that yield pseudo data - as simulated from the assumed model -
that matches the observed data, with the matching based on summary statistics.
If such statistics are sufficient, and if an arbitrarily small tolerance is
used in the matching, the selected draws can be used to produce a posterior
distribution that is exact up to simulation error; otherwise, an estimate of
the \textit{partial }posterior -\textbf{\ }defined as the density{\ of the
unknown parameters conditional on the }summary statistics - is the only
possible outcome.

The choice of statistics for use within ABC, in addition to techniques for
determining the matching criterion, are clearly of paramount importance, with
much recent research having been devoted to devising ways of ensuring that the
information content of the chosen set of statistics\textbf{\ }is maximized, in
some sense; e.g.\textbf{\ }Joyce and Marjoram (2008), Blum (2010) and
Fearnhead and Prangle (2012). In this vein, Drovandi \textit{et al.}
(2011),\textbf{\ }Creel and Kristensen (2015), Creel \textit{et al., }(2015)
and Drovandi \textit{et al.} (2015), produce statistics via an
\textit{auxiliary }model selected to approximate the features of the true data
generating process. This approach mimics, in a Bayesian framework, the
principle underlying the frequentist methods of indirect inference
(Gouri\'{e}roux \textit{et al.}, 1993) and efficient method of moments
(Gallant and Tauchen, 1996) using, as it does,\textbf{\ }the approximating
model to produce\textbf{\ }feasible inference about an intractable true model.
Whilst the price paid for the approximation in the frequentist setting is a
possible reduction in efficiency, the price paid in the Bayesian case is
posterior inference that is conditioned on statistics that are not sufficient
for the parameters of the true model, and which amounts to only partial
inference as a consequence.

Our paper continues in this spirit, but with focus {now }given to the
application of auxiliary model-based ABC methods in a general state space
model (SSM) framework and using the maximum of an auxiliary likelihood
function as the (vector) summary statistic.\textbf{\ }Drawing on recent
theoretical results on the properties of maximum likelihood estimation in
misspecified SSMs (Douc and Moulines, 2012) we provide a set of conditions
that ensures auxiliary likelihood-based ABC is\ \textit{Bayesian consistent}
in the state space setting, in the sense of producing draws that yield a
degenerate distribution at the true vector of static parameters in the (sample
size) limit. {The conditions for this {Bayesian consistency} result to hold
are cast explicitly in terms of auxiliary likelihood-based ABC, and exploit
the properties of the auxiliary (and, hence, misspecified) maximum likelihood
estimator (MLE) in the state space setting. As such, the results are both
distinct from, and complement, related asymptotic results in {Li and Fearnhead
(2018a,b)}{ and\ Frazier \textit{et al.} (2018) that pertain to the
application of ABC using generic summaries that are not explicitly defined
with reference to }an auxiliary likelihood function or a specific model
structure. }

The use of maximum likelihood to estimate the auxiliary parameters allows the
concept of asymptotic sufficiency to be invoked, thereby ensuring that - at
least for large samples - maximum information is extracted from the auxiliary
model in producing the summaries. {A} {selection criterion} based on the score
of the auxiliary likelihood - evaluated at the MLE computed from the observed
data - {is shown to yield }equivalent {draws to the criterion based on the}
MLE {itself, }for large enough sample sizes.\textbf{ }This equivalence is
satisfied in both the exactly- and over-identified cases, and implies that the
proximity to asymptotic sufficiency yielded by using the auxiliary MLE in an
ABC algorithm is replicated by the use of the auxiliary score. {ABC based on
the score of an auxiliary SSM likelihood is also proven to be Bayesian
consistent under regularity conditions. Given the enormous reduction in
computational cost afforded by the score approach (avoiding as it does the
need to optimize the auxiliary likelihood at each iteration of ABC) these
theoretical results are critically important for the application of ABC in
complex SSMs.}

Finally, we propose a numerical approach to\textbf{\ }circumvent the issue of
dimensionality that impacts on ABC techniques in multiple parameter settings.
(See Blum, 2010, Fearnhead and Prangle, 2012, {and} Nott \textit{et
al}\emph{.}, 2014). Specifically, we demonstrate numerically the improved
accuracy that can be achieved by matching individual parameters via the
corresponding scalar score of the integrated auxiliary likelihood, as an
alternative to matching on the multi-dimensional score statistic as suggested,
for example, in Drovandi \textit{et al.} (2015).

We illustrate the proposed method in three classes of stochastic volatility
{model} for {financial} asset returns. Two of the classes exemplify the case
where the transition densities in the state process have a representation that
is either challenging to embed within an exact algorithm or is unavailable
analytically. The third class of model illustrates the case where the
conditional density of returns given the latent volatility is unavailable.
Examples from all three classes are then explored numerically, in artificial
data scenarios. {In one particular example, in which the exact marginal
posteriors are accessible, the {{accuracy} of {auxiliary score-based} ABC {is
explored}, relative to a range of comparators }that includes }{two particle
{marginal }Metropolis Hastings (PMMH) {algorithms}.} Additionally, in the
supplementary material, we{\textbf{ }numerically verify Bayesian {consistency
{of the auxiliary score-based approach }for specific models} in each of the
three {stochastic volatility} classes. This being the first attempt made {to
verify} the {accuracy and asymptotic }validity of auxiliary likelihood-based
ABC techniques in such complex settings, the results augur well for the future
use of the method. }

The paper proceeds as follows. In Section \ref{abc} we briefly summarize the
basic principles of ABC as they would apply in a state space
framework.\textbf{\ }In Section \ref{Aux}, we then proceed to demonstrate the
theoretical\textbf{\ }properties of the auxiliary likelihood approach to ABC,
including sufficient conditions for Bayesian consistency to hold, in this
particular setting. The sense in which inference based on the auxiliary MLE is
replicated by inference based on the auxiliary score is also described. In
Section \ref{model_vol} we then consider the application of the auxiliary
likelihood approach in the non-linear state space setting, using the three
classes of {stochastic} volatility models for illustration.

Numerical accuracy of the proposed method, as applied to data generated
artificially from {one particular volatility model - the {continuous-time
Heston} (1993) square root model - }is then assessed in {Section
\ref{assess}}. Using a deterministic non-linear filtering technique (Ng \textit{et
al}., 2013), exact posteriors for the parameters in the {Heston model} are
attainable at an arbitrary level of numerical precision. The existence of
these, numerically, exact posteriors allows us to conduct a meaningful
comparison between the approximate posteriors yielded by ABC and the
posteriors produced by PMMH methods, which are exact up to simulation error.
Specifically, in the context of this model, we compare the accuracy of
posterior estimates across 18 different competitors: 16 ABC-based comparators,
including variants of our {auxiliary }score-based approach, and using various
dimension reduction techniques - regression adjustment, the approach of
Fearnhead and Prangle (2012), and our proposed integrated auxiliary likelihood
method; plus two PMMH comparators, one based on the bootstrap particle filter
({Andrieu \textit{et al., }2010; Pitt \textit{et al., }}2012), and the other
using an ABC filtering step within the MCMC chain ({Dean \textit{et al., }%
}2014; Calvet and Czellar, 2015; Jasra, 2015). {A key result is that a
particular auxiliary score-based ABC estimate of the exact marginals is {the
most} accurate {method }overall, {including in comparison with }both PMMH
methods.} Furthermore, {in the }supplementary material, we present
additional numerical evidence supporting Bayesian consistency for the
auxiliary-score based ABC approach in all three {volatility models},
while, in contrast, evidence for the consistency of various other ABC methods
is mixed. {An empirical illustration of the score-based method in a setting in
which exact inference is essentially infeasible follows in Section
\ref{empiric}, whilst }Section \ref{end} concludes. Technical proofs,
additional numerical results, certain computational details and the computer
code used to produce the numerical results are provided in {the supplementary
material}.

\section{ABC in state space models\label{abc}}

\subsection{An outline of ABC}

The aim of ABC is to produce draws from an approximation to the posterior
distribution of a vector of unknowns, $\mathbf{\theta}$, given the
$T$-dimensional vector of observed data $\mathbf{y}=(y_{1},...,y_{T})^{\prime
}$,$$p(\mathbf{\theta|y})\propto
p(\mathbf{y|\theta})p(\mathbf{\theta}),$$ in the canonical case where both the
prior, $p(\mathbf{\theta})$, and the likelihood, $p(\mathbf{y|\theta})$, can
be simulated. These draws are used, in turn, to approximate posterior
quantities of interest, including marginal posterior moments, marginal
posterior distributions and predictive distributions. The simplest
(accept/reject) form of the algorithm (Tavar\'{e}\textit{\ et al.}%
,\textit{\ }1997; Pritchard, 1999) proceeds as per Algorithm 1.
\begin{algorithm}
\caption{ABC accept/reject algorithm}\label{ABC}
\begin{algorithmic}[1]
\State Simulate $\mathbf{\theta }^{i}$, $i=1,2,...,N$, from $p(\mathbf{\theta })$
\State Simulate $\mathbf{z}^{i}=(z_{1}^{i},z_{2}^{i},...,z_{T}^{i})^{\prime }$, $i=1,2,...,N$, from the likelihood, $p(\mathbf{.|\theta }^{i})$
\State Select $\mathbf{\theta }^{i}$ such that:%
\begin{equation}
d\{\mathbf{\eta }(\mathbf{y}),\mathbf{\eta }(\mathbf{z}^{i})\}\leq
\varepsilon ,  \label{distance}
\end{equation}%
where $\mathbf{\eta} (\mathbf{.})$ is a (vector) statistic, $d\{.\}$ is a distance
criterion, and, given $N$, the tolerance level $\varepsilon $ is chosen to be small.
\end{algorithmic}
\end{algorithm}The algorithm thus samples $\mathbf{\theta}$ and $\mathbf{z}$
from the joint\textit{\ }posterior:%
\[
p_{\varepsilon}(\mathbf{\theta},\mathbf{z|\eta(y)})=\frac{p(\mathbf{\theta
})p(\mathbf{z|\theta})\mathbb{I}_{\varepsilon}[\mathbf{z}]}{\textstyle\int%
_{\mathbf{\Theta}}\int_{\mathbf{z}}p(\mathbf{\theta})p(\mathbf{z|\theta
})\mathbb{I}_{\varepsilon}[\mathbf{z}]d\mathbf{z}d\mathbf{\theta}},
\]
where $\mathbb{I}_{\varepsilon}[\mathbf{z}]$:=$\mathbb{I}[d\{\mathbf{\eta
}(\mathbf{y}),\mathbf{\eta}(\mathbf{z})\}\leq\varepsilon]$ is one if
$d\left\{  \mathbf{\eta}(\mathbf{y}),\mathbf{\eta}(\mathbf{z})\right\}
\leq\varepsilon$ and zero else. Clearly, when $\mathbf{\eta}(\mathbf{\cdot})$
is sufficient and $\varepsilon$ small,%
\begin{equation}
p_{\varepsilon}(\mathbf{\theta|\eta(y)})=\textstyle\int_{\mathbf{z}%
}p_{\varepsilon}(\mathbf{\theta},\mathbf{z|\eta(y)})d\mathbf{z}%
\label{abc_post}%
\end{equation}
is a good approximation to the exact posterior, $p(\mathbf{\theta|y})$, and
draws from $p_{\varepsilon}(\mathbf{\theta},\mathbf{z|\eta(y)})$ can be used
to estimate features of that exact posterior. In practice however, the
complexity of the models to which ABC is applied, including in the state space
setting, {is such that} that {a {low-dimensional} set of sufficient statistics
is unavailable}.\textbf{\ }Hence, as $\varepsilon\rightarrow0$ the draws can
be used to estimate features of $p(\mathbf{\theta|\eta}(\mathbf{y}))$ only$.$

Adaptations of the basic rejection scheme have involved post-sampling
corrections of the draws using kernel methods (e.g. Beaumont \textit{et al.},
2002; Blum, 2010), or the insertion of{ MCMC} and/or sequential Monte Carlo
(SMC) steps (Marjoram \textit{et al.}, 2003; Sisson \textit{et al.,} 2007;
Beaumont \textit{et al.}, 2009; Wegmann \textit{et al.}, 2009), to improve the
accuracy with which $p(\mathbf{\theta|\eta}(\mathbf{y}))$ is estimated, for
any given number of draws. Focus is also given to choosing $\mathbf{\eta}(.)$
and/or $d\{.\}$ so as to render $p(\mathbf{\theta|\eta}(\mathbf{y}))$ a closer
match to $p(\mathbf{\theta|y})$, in some sense; see Joyce and Marjoram (2008),
Wegmann \textit{et al.}, Blum (2010) and Fearnhead and Prangle (2012). In the
latter vein, Drovandi \textit{et al.} (2011) were the first to argue, in the
context of a specific biological model, that the use of $\mathbf{\eta}(.)$
comprised of the MLEs of the parameters of a well-chosen approximating model,
may yield posterior inference that is conditioned on a large portion of the
information in the data and, hence, be close to exact inference based on
$p(\mathbf{\theta|y})$. It is the spirit of this approach that informs the
current paper, but with our attention given to rendering the approach feasible
in a general state space framework that encompasses a large number of the
models that are of interest to practitioners.

\subsection{Auxiliary likelihood-based ABC in SSMs: Basic approach\label{ssm}}

The stochastic process $\{y_{t}\}_{t\geq0}$ represents a stationary ergodic
process taking values in a measure space $(Y,\mathcal{F}_{y})$, with
$\mathcal{F}_{y}$ a Borel $\sigma$-field, specified according to an SSM that
depends on an unobserved state process $\{x_{t}\}_{t\geq0}$, taking values in
a measure space $(X,\mathcal{F}_{x})$, with $\mathcal{F}_{x}$ a Borel $\sigma
$-field. {Whilst not necessary for what follows, we choose to illustrate our
approach in the case where both }$x_{t}$ {and }$y_{t}$ {are scalars. Any
extension to the vector case would incur a concurrent increase in the
dimension of the unknown parameters on which the process generating }$x_{t}$
{and }$y_{t}$ {depends and an associated need to manage the impact of
dimension on the ABC algorithm. However, no other aspect of what we propose
would be materially altered.}

The SSM is parameterized by {a vector of unknown parameters} $\mathbf{\phi}%
\in\Phi\subset\mathbb{R}^{d_{\phi}}$,\textbf{ }with the parameter space $\Phi$
assumed to be compact, and for each $\mathbf{\phi}$, the state and observed
sequences are generated according to the following measurement and state
equations:
\begin{align}
y_{t}  &  =b(x_{t},w_{t},\mathbf{\phi)} ,\label{d1}\\
x_{t}  &  =G_{\mathbf{\phi}}(x_{t-1})+\Sigma_{\mathbf{\phi}}(x_{t-1})v_{t},
\label{d2}%
\end{align}
where\textbf{\ }$\{w_{t},v_{t}\}_{t\geq0}$ are independent sequences of
$i.i.d.$ random variables, $b(\cdot),\Sigma_{\mathbf{\phi}}(\cdot
),G_{\mathbf{\phi}}(\cdot)$ are known, potentially nonlinear functions
depending on $\mathbf{\phi}\in\Phi$, and where $\Sigma_{\mathbf{\phi}}%
(\cdot)>0$ for all $\mathbf{\phi}\in\Phi$. For each $\mathbf{\phi}\in\Phi$, we
assume that equation (\ref{d2}) defines a transition density $p(x_{t}%
|x_{t-1},\mathbf{\phi})$ and that equation (\ref{d1}) gives rise to the
conditional density of $y_{t}$. This\textbf{\ }allows us to state
the{\ measurement and transition densities }respectively{\textbf{\ }as:}%
\begin{align}
&  p(y_{t}|x_{t},\mathbf{\phi}) ,\label{meas_gen}\\
&  p(x_{t}|x_{t-1},\mathbf{\phi}) , \label{state_gen}%
\end{align}
{for }$t=1,2,...,T$, {with }$x_{0}$ {assumed to follow the stationary
distribution of }$x_{t}.$ Throughout the remainder, we denote the `true value'
generating $\{y_{t}\}_{t\geq0}$ by $\mathbf{\phi}_{0}\in\Phi$, and denote by
$\mathbb{P}$ and $\mathbb{E}$ the law and expectation of the stationary SSM
associated with $\mathbf{\phi}_{0}$.

The aim of the current paper is to use ABC principles to conduct inference
about (\ref{meas_gen}) and (\ref{state_gen}) through $\mathbf{\phi}$. Our
particular focus is situations where at least one of (\ref{meas_gen}) or
(\ref{state_gen}) is analytically unavailable, or computationally challenging,
such that exact MCMC- or SMC-based techniques are infeasible or, at the very
least, computationally burdensome. Three such classes of examples are later
explored in detail, with all examples related to the modelling of stochastic
volatility for financial returns, and with one example highlighting the case
of a continuous-time volatility process.

ABC methods can be implemented within these types of settings so long as
simulation from (\ref{meas_gen}) and (\ref{state_gen}) is straightforward and
appropriate\textbf{\ }`summaries' of the data are available. We conduct
ABC-based inference by relying on the structure of the SSM in (\ref{d1}) and
(\ref{d2}) to generate a simplified version of the SSM, which we then use to
produce informative summary measures for use in ABC. Specifically, we consider
a simplified and, hence, misspecified\textbf{\ }version of equations
(\ref{d1}) and (\ref{d2}), where
\begin{align}
y_{t}  &  =a(x_{t},\epsilon_{t},\mathbf{\beta)} ,\label{miss1}\\
x_{t}  &  =H_{\mathbf{\beta}}(x_{t-1})+S_{\mathbf{\beta}}(x_{t-1})e_{t},
\label{miss2}%
\end{align}
with $\{\epsilon_{t},e_{t}\}_{t\geq0}$ independent sequences of\textbf{\ }%
$i.i.d.$\textbf{\ }random variables with well-behaved densities;
$a(\cdot),S_{\mathbf{\beta}}(\cdot),H_{\mathbf{\beta}}(\cdot)$ known functions
of unknown parameters $\mathbf{\beta}\in\mathcal{B}\subset\mathbb{R}%
^{d_{\beta}}$; and $S_{\mathbf{\beta}}\left(  \cdot\right)  >0$ for all
$\mathbf{\beta}$. Together, we assume this specification ensures that
$\{x_{t}\}_{t\geq0}$ takes values in the measure space $(X,\mathcal{F}_{x})$
and leads to a known transition kernel $Q_{\beta}:X\times\mathcal{A}%
\times\mathcal{B}\rightarrow\lbrack0,1]$, $\mathcal{A}\in\mathcal{F}_{x}$,
which admits the known state-transition density $q_{\beta}(\cdot
,\cdot):X\times X\times\mathcal{B}\rightarrow\mathbb{R}_{+}$, and known
conditional density $g_{\beta}:X\times Y\times\mathcal{B}\rightarrow
\mathbb{R}_{+}$. That is, equations \eqref{miss1} and \eqref{miss2} yield the
state density $q_{\beta}(x_{t},x_{t-1})$ and measurement density $g_{\beta
}(y_{t},x_{t})$ for the auxiliary model, with both $q_{\beta}(\cdot,\cdot)$
and $g_{\beta}(\cdot,\cdot)$ analytically tractable.

Defining the parametric family of the above misspecified SSM
as$$\mathcal{G}:=\{(q_{\beta
}(x,x^{^{\prime}}),g_{\beta}(y,x)):\mathbf{\beta}\in\mathcal{B},\;y\in
Y,\;x,x^{\prime}\in X\},$$ we maintain that there is no reason to assume
$\mathbb{P}\in\mathcal{G}$. However, even if $\mathbb{P}\notin\mathcal{G}$, it
will generally be the case that a well-chosen $\mathcal{G}$ is capable of
capturing many of the features associated with the {data generating process}
in equations (\ref{d1}) and (\ref{d2}). To this end, and in the spirit of
indirect inference, we obtain summary statistics for ABC using the
quasi-likelihood associated with the parametric family $\mathcal{G}$. Such a
strategy requires defining the quasi-likelihood associated with the
misspecified SSM, which, following Gouri\'{e}roux \textit{et al}. (1993),
amongst others, is hereafter referred to as the auxiliary likelihood. Defining
$\chi(\cdot)$ to be an initial probability measure on $(X,\mathcal{F}_{x})$,
we state the auxiliary likelihood for inference on $\mathbf{\beta}$ as
\[
\ell(\mathbf{y};\mathbf{\beta})=\int\cdots\int\chi(dx_{0})g_{\beta}(y_{0}%
,{x}_{0})\prod_{p=1}^{T}Q_{\beta}(x_{p-1},dx_{p})g_{\beta}(y_{p},x_{p}).
\]
From observations $\mathbf{y}$, the auxiliary\textbf{\ }MLE can then be
obtained as
\begin{equation}
\widehat{\mathbf{\beta}}(\mathbf{y)}=\arg\max_{\beta\in\mathcal{B}}%
L_{a}(\mathbf{y};\mathbf{\beta});\;L_{a}(\mathbf{y};\mathbf{\beta})=\log
(\ell(\mathbf{y};\mathbf{\beta})).\label{ss_like}%
\end{equation}
Given $\mathbf{\eta(y)}=\widehat{\mathbf{\beta}}(\mathbf{y)}$, ABC can
then\textbf{\ }proceed via Algorithm 1.

We note that, in the above setting, the full set of unknowns constitutes the
augmented vector $\mathbf{\theta}=(\mathbf{\phi}^{\prime}\mathbf{,x}%
_{c}^{\prime})^{\prime}$ where, in the case when $x_{t}$ evolves in continuous
time, $\mathbf{x}_{c}$ represents the infinite-dimensional vector comprising
the continuum of unobserved states over the sample period. However, to fix
ideas, we define $\mathbf{\theta}=(\mathbf{\phi}^{\prime}\mathbf{,x}^{\prime
})^{\prime},$ where $\mathbf{x}=(x_{1},x_{2},...,x_{T})^{\prime}$ is the
$T$-dimensional vector comprising the time $t$ states for the $T$ observation
periods in the sample.\footnote{For example, in a continuous-time stochastic
volatility model such values may be interpreted as end-of-day volatilities.}
Implementation of {Algorithm 1} thus involves simulating $\mathbf{\phi}$ from
the prior $p(\mathbf{\phi})$, followed by simulation of $x_{t}$ via the
process for the state, conditional on the draw of $\mathbf{\phi}$, and
subsequent simulation of artificial data $z_{t}$ conditional on the draws of
$\mathbf{\phi}$ and the state variable. Crucially, our attention is given to
inference about $\mathbf{\phi}$\ only; hence, only draws of $\mathbf{\phi}%
$\ are retained (via the selection criterion) and those draws used to produce
an estimate of the marginal posterior, $p(\mathbf{\phi|y})$. That is, from
this point onward, when we reference a vector of summary statistics,
$\mathbf{\eta(y)}$, for instance, $\mathbf{\eta(y)}=\widehat{\mathbf{\beta}%
}(\mathbf{y)},$ it is the information content of that vector with respect to
$\mathbf{\phi}$\ that is of importance, and the asymptotic behaviour of
$p_{\varepsilon}(\mathbf{\phi|\eta(y)})$\ with reference to the true
$\mathbf{\phi}_{0}$ that is under question. Similarly, in the numerical
illustration in Section \ref{sq_numerics}, it is the proximity of {any particular}
(kernel-based estimate of)\ $p_{\varepsilon}(\mathbf{\phi|\eta(y)})$ explored
therein to the exact\textbf{\ }$p(\mathbf{\phi|y})$\textbf{\ }that is
documented.\textbf{\ }We comment briefly on state inference in Section
\ref{end}.

\section{Auxiliary likelihood-based ABC in SSMs: Theory and
computation\label{Aux}}

\subsection{`Approximate' asymptotic sufficiency\label{TheoryS}}

ABC is predicated on the use of `informative' summaries in its implementation,
with a vector of sufficient statistics being the only form of summary that
replicates the information content of the full sample, and with {the
Pitman-Koopman-Darmois theorem establishing that }sufficiency ({via a set of
statistics that is lower in dimension than the full sample}) is attainable
only for distributions that are members of the exponential family. For the
general SSM described by (\ref{meas_gen}) and (\ref{state_gen}) for any $t$ -
and with our particular focus being cases where either density does not have
an analytical representation - the joint distribution of $\mathbf{y}$ will,
almost by default, not be in the exponential family, and sufficiency reduction
will therefore not be feasible.\footnote{Even the simplest SSMs, with all
components available, generate moving average-like dependence in the data. The
linear Gaussian SSM is the leading case, and for which simple computations
lead to an analytical link between the signal-to-noise ratio and the lack of
sufficiency associated with any finite set of statistics calculated from the
observations. The crux of the problem is that information in the sample does
not `accumulate' in the way required for reduction to a sufficient set of
statistics of dimension smaller than $T$ to be feasible (see, for example,
Anderson, 1958, Chp. 6). The essence of this problem would characterize any
SSM nested in (\ref{meas_gen}) and (\ref{state_gen}), simply due to the
presence of measurement error.}

On the other hand, {limiting} Gaussianity of the MLE for the parameters of
(\ref{meas_gen}) and (\ref{state_gen}) {implies that, under regularity, }the
MLE\ (asymptotically) satisfies the factorization theorem and is thereby
asymptotically sufficient for the parameters of that model. Denoting the
log-likelihood function by $L({\mathbf{y;\phi}})$, maximizing
$L({\mathbf{y;\phi}})$ with respect to ${\mathbf{\phi}}$ yields
$\widehat{{\mathbf{\phi}}}$, which could, \textit{in principle}, be used to
define ${\mathbf{\eta}}({\mathbf{.}})$ in an ABC algorithm. For large enough
$T$ (and for $\varepsilon\rightarrow0$) the algorithm would thus produce draws
from the exact posterior. Indeed, in arguments that mirror those adopted by
Gouri\'{e}roux \textit{et al.} (1993) and Gallant and Tauchen (1996) for the
indirect inference and efficient method of moments estimators respectively, if
${\mathbf{\eta}}({\mathbf{.}})$ is chosen to be the MLE of an auxiliary model
that {nests (or `smoothly embeds'}) the true model in some well-defined way,
asymptotic sufficiency for the true parameters\textbf{\ }will still be
achieved; see also Gouri\'{e}roux and Monfort (1995) on this point.

Of course, if the SSM in question is such that the exact likelihood is
accessible, the model is likely to be tractable enough to preclude the need
for treatment via ABC, with the primary goal of this paper being the
presentation of ABC methods in SSMs for which exact methods are essentially
infeasible. Further, the quest for asymptotic sufficiency via a
\textit{nesting} auxiliary model conflicts with the quest for an accurate
non-parametric estimate of the posterior using the ABC draws, given that the
dimension of the parameter set in the auxiliary model is, by construction,
likely to be large. Hence, in practice, the appropriate goal in using the
auxiliary likelihood approach to ABC in the SSM context is to define, via
(\ref{miss1}) and (\ref{miss2}), a sensible \textit{parsimonious}
approximation to the true model\textbf{\ }in (\ref{meas_gen}) and
(\ref{state_gen}), for which the associated likelihood function can be
evaluated with computational ease and speed. Heuristically, if the
approximating model is `accurate enough' as a representation of the true
model, such an approach will yield, via the ABC algorithm, an estimate of the
posterior distribution that is conditioned on a statistic that is `close to'
being asymptotically sufficient for\textbf{\ }$\mathbf{\phi}$. We certainly
make no attempt in this paper to formalize this statement in any way.
Nevertheless, we do view the notion of asymptotic sufficiency of the auxiliary
MLE as being a intuitively compelling characteristic of the auxiliary
likelihood-based approach to ABC, and the numerical results presented later
provide some support for its importance in practice. More critically, however,
pursuing the auxiliary likelihood route enables us to draw on regularity as it
pertains to likelihood functions, and maximization thereof, to prove the
Bayesian consistency of the resultant ABC posterior and, hence, the baseline
accuracy of the inferences produced via this route.

\subsection{Consistency of auxiliary likelihood-based ABC\label{Theory}}

For a given choice of auxiliary model in (\ref{miss1}) and (\ref{miss2}), with
parameters ${\mathbf{\beta}}\in{\mathcal{B}}\subset{\mathbb{R}}^{d_{\beta}}$,
$d_{\beta}\geq d_{\phi}$, and sample log-likelihood function $L_{a}%
({\mathbf{y}};{\mathbf{\beta}})$ defined in (\ref{ss_like}), ABC can use as
summary statistics for inference on $\mathbf{\phi}$ the maximizers of
$L_{a}({\cdot};{\mathbf{\beta}})$, based on $\mathbf{y}$ and $\mathbf{z(\phi
}^{i})$, which we represent respectively by
\[
\widehat{{\mathbf{\beta}}}({\mathbf{y}})=\arg\max_{{\mathbf{\beta}}%
\in{\mathcal{B}}}L_{a}({\mathbf{y}};{\mathbf{\beta}})\text{ and }%
\widehat{{\mathbf{\beta}}}(\mathbf{z(\phi}^{i}))=\arg\max_{{\mathbf{\beta}}%
\in{\mathcal{B}}}L_{a}(\mathbf{z(\phi}^{i});{\mathbf{\beta}}).
\]
Herein,\textbf{\ }$\mathbf{z(\phi}^{i})$ is the\textbf{\ }$ith$\textbf{\ }%
vector of pseudo data, with the dependence of\textbf{\ }$\mathbf{z(\phi}^{i})$
on the\textbf{\ }$ith$\textbf{\ }random draw\textbf{\ }$\mathbf{\phi}^{i}%
$\textbf{\ }from the prior\textbf{\ }$p({\mathbf{\phi)}}$ {made explicit in
the notation}. {Using} $\mathbf{\eta}(\mathbf{y})=\widehat{{\mathbf{\beta}}%
}({\mathbf{y}})$ and$\;{\mathbf{\eta}}(\mathbf{z(\phi}^{i}
))=\widehat{{\mathbf{\beta}}}(\mathbf{z(\phi}^{i}))$ as summary statistics, we
can take as the distance criterion in \eqref{distance},
\begin{equation}
d\{{\mathbf{\eta}}({\mathbf{y}}),{\mathbf{\eta}}(\mathbf{z(\phi}^{i}%
))\}=\sqrt{\left[  \widehat{{\mathbf{\beta}}}({\mathbf{y)-}}%
\widehat{{\mathbf{\beta}}}(\mathbf{z(\phi}^{i}){\mathbf{)}}\right]  ^{\prime
}{\mathbf{\Omega}}\left[  \widehat{{\mathbf{\beta}}}({\mathbf{y)-}%
}\widehat{{\mathbf{\beta}}}(\mathbf{z(\phi}^{i}){\mathbf{)}}\right]  },
\label{dist_mle}%
\end{equation}
where ${\mathbf{\Omega}}$ is some positive definite matrix.

As noted above, {the use of a parsimonious (non-nesting) auxiliary model means
that asymptotic }sufficiency {for }$\mathbf{\phi}$ {is not attainable}. As
such, beyond adhering to the principle of choosing an accurate approximating
model and thereby attaining a summary statistic that is `not far from' being
asymptotically sufficient,\textbf{ }we require some guarantee that
$p_{\varepsilon}(\mathbf{\phi}|\mathbf{\eta}(\mathbf{y}))$ yields reasonable,
and statistically valid, inference in the complex SSMs that are our focus. To
this end, we establish conditions under which {auxiliary likelihood-based }ABC
attains a relatively weak - but no less important - form of validity, namely
Bayesian consistency. Under such conditions the investigator can be assured
that, at the very least, with a large enough sample size the ABC posterior
will concentrate on the true parameter vector and provide valid inference in
that sense.

In the ABC setting, Bayesian consistency requires that as $T\rightarrow\infty$
and\textbf{\ }$\varepsilon\rightarrow0$, the estimated posterior based on the
selected draws from $p_{\varepsilon}(\mathbf{\phi|\eta(y)})$ concentrates
around\textbf{\ }the true parameter value generating the data; see {Li and
Fearnhead (2018a,b)} and\textbf{\ }Frazier \textit{et al.} (2018) for related
discussion on asymptotic concepts as they pertain to ABC. With a slight abuse
of terminology, from this point onwards we denote the `ABC posterior'
by\textbf{\ }$p_{\varepsilon}({\mathbf{\phi}}|{\mathbf{\eta}}({\mathbf{y}}))$,
recognizing that the quantity produced via ABC is actually the kernel-based
density estimate constructed from a given number of draws, $N$, from
$p_{\varepsilon}(\mathbf{\phi}|\mathbf{\eta}(\mathbf{y}))$ as defined in
(\ref{abc_post}).

To understand the intuition underlying Bayesian consistency of ABC based on
${\mathbf{\eta}}({\mathbf{y}})=\widehat{{\mathbf{\beta}}}({\mathbf{y}})$,
first define $Z\subseteq Y$ to be the space of simulated data $\mathbf{z}%
(\mathbf{\phi})$, generated according to the\textbf{\ }probability measure
$P_{z}^{\phi}$, and denote the prior measure of a set $A\subset{\Phi}$ by
$\Pi(A)$. We also make it explicit from this point onwards that Bayesian
consistency depends on simultaneous asymptotics regarding $T$ and
$\varepsilon.$ To formalize this we consider $\varepsilon$ as a $T$-dependent
sequence, denoted by $\varepsilon_{T}$, where $\varepsilon_{T}$ $\rightarrow0$
as $T$\ $\rightarrow\infty.$

Heuristically, Bayesian consistency of ABC would then follow from the
following sequence of arguments. First, as $T\rightarrow\infty,$ the criterion
in (\ref{dist_mle}) should satisfy (uniformly)
\begin{equation}
d\{{\mathbf{\eta}}({\mathbf{y}}),{\mathbf{\eta}}(\mathbf{z(\phi}%
^{i}))\}\xrightarrow {P}\sqrt{\left[  \mathbf{\beta}_{0}-{\mathbf{b}%
}({\mathbf{\phi}}^{i}{\mathbf{)}}\right]  ^{\prime}{\mathbf{\Omega}}\left[
\mathbf{\beta}_{0}-{\mathbf{b}}({\mathbf{\phi}}^{i}{\mathbf{)}}\right]  },
\label{conv1}%
\end{equation}
where \textquotedblleft$\xrightarrow{P}$" denotes convergence in probability,
and where%
\[
\mathbf{\beta}_{0}=\arg\max_{{\mathbf{\beta}}\in{\mathcal{B}}}\left\{
\text{plim}_{T\rightarrow\infty}L_{a}(\mathbf{y};\mathbf{\beta})/T\right\}
;\text{ }{\mathbf{b}}({\mathbf{\phi}}^{i}{\mathbf{)=}}\arg\max_{{\mathbf{\beta
}}\in{\mathcal{B}}}\left\{  \text{plim}_{T\rightarrow\infty}L_{a}%
(\mathbf{z(\phi}^{i});\mathbf{\beta})/T\right\}  ,
\]
{where } $\text{plim}_{T\rightarrow\infty}X_T$ {denotes the probability limit of }$X_T$. Secondly,\textbf{\ }$\mathbf{\phi}^{i}=$\textbf{\ }%
$\mathbf{\phi}_{0}$ should be the only value that satisfies\textbf{\ }%
$\mathbf{\beta}_{0}=\mathbf{b}(\mathbf{\phi}^{i})$ and, as a consequence, the
only value that satisfies%
\begin{equation}
d\{\mathbf{\beta}_{0},\mathbf{b}({\mathbf{\phi}}^{i})\}=\sqrt{\left[
\mathbf{\beta}_{0}-{\mathbf{b}}({\mathbf{\phi}}^{i}{\mathbf{)}}\right]
^{\prime}{\mathbf{\Omega}}\left[  \mathbf{\beta}_{0}-{\mathbf{b}%
}({\mathbf{\phi}}^{i}{\mathbf{)}}\right]  }=0. \label{limit_distance}%
\end{equation}
Hence, as $T\rightarrow\infty$, for any $\varepsilon_{T}>0$ such that
$\Pi\lbrack\{\mathbf{\phi}^{i}\in{\Phi}:d\{\mathbf{\beta}_{0},\mathbf{b}%
({\mathbf{\phi}}^{i})\}\leq\varepsilon_{T}\}]>0,$ the only value of
$\mathbf{\phi}^{i}$ satisfying $d\{\mathbf{\eta}(\mathbf{y}),\mathbf{\eta
}(\mathbf{z(\phi}^{i}))\}\leq\varepsilon_{T}$ for all $\varepsilon_{T}$ is
${\mathbf{\phi}}^{i}={\mathbf{\phi}}_{0}$; therefore, if
$\widehat{{\mathbf{\beta}}}({\mathbf{y}})$ is well-behaved, as $T\rightarrow
\infty,$ $\varepsilon_{T}\rightarrow0$, the ABC algorithm will only select
draws arbitrarily close to ${\mathbf{\phi}}_{0}$. Put formally, the ABC
posterior will be Bayesian consistent if, for any $\delta>0$ and $A_{\delta
}({\mathbf{\phi}}_{0}):=\{{\mathbf{\phi}}\in{{\Phi}}:d\left\{  {\mathbf{\phi}%
},{\mathbf{\phi}}_{0}\right\}  >\delta\}$,%
\begin{equation}
\int_{A_{\delta}({\mathbf{\phi}}_{0})}p_{\varepsilon}({\mathbf{\phi}%
}|{\mathbf{\eta}}({\mathbf{y}}))d{\mathbf{\phi}}=\int_{A_{\delta
}({\mathbf{\phi}}_{0})}\frac{\int_{Z}\mathbb{I}\left[
d\{\widehat{{\mathbf{\beta}}}(\mathbf{y}),\widehat{{\mathbf{\beta}}%
}(\mathbf{z}({\mathbf{\phi}}))\}\leq\varepsilon_{T}\right]  P_{z}^{\phi
}(d\mathbf{z)}\Pi(d\mathbf{\phi})}{\int_{\Phi}\int_{Z}\mathbb{I}\left[
d\{\widehat{{\mathbf{\beta}}}(\mathbf{y}),\widehat{{\mathbf{\beta}}%
}(\mathbf{z}({\mathbf{\phi}}))\}\leq\varepsilon_{T}\right]  P_{z}^{\phi
}(d\mathbf{z)}\Pi(d\mathbf{\phi})}=o_{P}(1), \label{bc_1}%
\end{equation}
as $T\rightarrow\infty$ and $\varepsilon_{T}\rightarrow0$. Sufficient
conditions needed to demonstrate the convergence in \eqref{bc_1} can be split
into two sets: {the first }controls the convergence of sample quantities; the
second set {comprises} identification conditions. {Let $\ln^{+}(x)=\max
\{0,\ln(x)\}$ and $\ln^{-}(x)=-\min\{0,\ln(x)\}$.}

\noindent\textbf{Assumption A:}

\noindent\textbf{(A1)} The parameter spaces $\mathcal{B}\subset{\mathbb{R}%
}^{d_{\beta}}$ and ${{\Phi}}\subset{\mathbb{R}}^{d_{\phi}}$ are compact.

\noindent\textbf{(A2)} For any $\mathbf{\phi}\in{\Phi}$, $z_{t}(\mathbf{\phi
})\in Z\subseteq Y$, $\{z_{t}(\mathbf{\phi}),x_{t}(\mathbf{\phi})\}_{t=1}^{T}$
is a stationary and ergodic process, with $(z_{0}(\mathbf{\phi}),x_{0}%
(\mathbf{\phi}))$ drawn in the stationary law.

\noindent\textbf{(A3)} For $(x,x^{\prime},{\mathbf{\beta}})\mapsto q_{\beta
}(x,x^{\prime})$ the density of the Markov transition kernel associated with
the auxiliary model satisfies the following:

\hspace{.1cm}(A3.1) $(x,x^{\prime},{\mathbf{\beta}})\mapsto q_{\beta
}(x,x^{\prime})$ is a positive continuous function on $X\times X\times
\mathcal{B}$.

\hspace{.1cm}(A3.2) $\sup_{{\beta}\in\mathcal{B}}\sup_{(x,x^{\prime})\in
X\times X}q_{\beta}(x,x^{\prime})<\infty$.

\noindent\textbf{(A4)} The conditional density $(y,x,\mathbf{\beta})\mapsto$
$g_{\beta}(y,x)$ associated with the auxiliary model satisfies the following conditions:

\hspace{.1cm}(A4.1) For each $(x,y)\in X\times Y$, $(y,x,\mathbf{\beta
})\mapsto g_{\beta}(y,x)$ is positive and continuous on $Y\times
X\times\mathcal{B}$.

\hspace{.1cm}(A4.2) For any $\mathcal{K}\subset Y$, compact, and any ${\beta
}\in\mathcal{B}$, $\lim_{|x|\rightarrow\infty}\sup_{y\in\mathcal{K}}%
\frac{g_{\beta}(y,x)}{\sup_{x^{\prime}\in X}g_{\beta}(y,x^{\prime})}=0.$

\hspace{.1cm}(A4.3) For $z_{0}({\phi})\in Y$ as in \textbf{(A2)},
$\mathbb{E}_{{\phi}}\left[  \ln^{+}\sup_{\beta\in\mathcal{B}}\sup_{x\in
X}g_{\beta}(z_{0}({\phi}),x)\right]  <\infty.$

\hspace{.1cm}(A4.4) There exists a compact subset $\mathcal{D}\subset X$ such
that, for $z_{0}({\phi})\in Y$ as in \textbf{(A2)}, $\mathbb{E}_{{\phi}%
}\left[  \ln^{-}\inf_{{\beta}\in\mathcal{B}}\inf_{x\in D}g_{\beta}(z_{0}%
({\phi}),x)\right]  <\infty.$

\noindent\textbf{(A5)} $L_{\infty}(\mathbf{\phi}^{i}\mathbf{;\beta}):=$
$\text{plim}_{T\rightarrow\infty}(1/T)L_{a}(\mathbf{z(\phi}^{i});\mathbf{\beta
})$ has unique maximum ${\mathbf{b}}(\mathbf{\phi}^{i})=\arg\max
_{{\mathbf{\beta}}\in{\mathcal{B}}}L_{\infty}(\mathbf{\phi}^{i}\mathbf{;\beta
})$, where\textbf{\ }$\mathbf{\beta}_{0}={\mathbf{b}}(\mathbf{\phi}_{0}%
)=\arg\max_{{\mathbf{\beta}}\in{\mathcal{B}}}L_{\infty}(\mathbf{\phi}%
_{0}\mathbf{;\beta}).$ \bigskip

\noindent\textbf{Assumption I:}

\noindent\textbf{(I1)} The prior $p(\mathbf{\phi})$ is absolutely continuous
with respect to the Lebesgue measure and satisfies $p(\mathbf{\phi}_{0})>0$.

\noindent\textbf{(I2)} The mapping ${\mathbf{\phi}}\mapsto{\mathbf{b}%
}({\mathbf{\phi}})$ is continuous and satisfies $\mathbf{\beta}_{0}%
=\mathbf{b(\phi)}\iff\mathbf{\phi}=\mathbf{\phi}_{0}$.\medskip

\noindent\textbf{(I3)} For any $\mathbf{\phi}\in\Phi$, there exist constants
$\kappa,C,u_{0}>0$ such that, for some sequence $v_{T}\rightarrow\infty$ and
all $0<u<u_{0}v_{T}$, for $\left\Vert \cdot\right\Vert $ the Euclidean norm
\[
P_{z}^{\phi}\left[  \left\|  \widehat{{\mathbf{\beta}}}(\mathbf{z}%
({\mathbf{\phi}}))-\mathbf{b}(\mathbf{\phi})\right\|  >u\right]  \leq C({\phi
})u^{-\kappa}v_{T}^{-\kappa},\text{ and }\int_{\Phi}C(\mathbf{\phi}%
)\Pi(d\mathbf{\phi})<\infty.
\]

\medskip

\noindent\textbf{Remark 1: }Under correct specification of the model
generating the data $\mathbf{y,}$ Assumptions \textbf{(A1)}-\textbf{(A5)
}ensure that $\sup_{\mathbf{\beta}\in\mathcal{B}}|(1/T)L_{a}(\mathbf{y}%
;\mathbf{\beta})-L_{\infty}(\mathbf{\phi}_{0}\mathbf{;\beta})|=o_{P}(1),$ for
$L_{\infty}(\mathbf{\phi}_{0}\mathbf{;\beta})$ defined in \textbf{(A5)}%
,\textbf{\ }and that $\Vert\widehat{{\mathbf{\beta}}}({\mathbf{y}%
})-\mathbf{\beta}_{0}\Vert=o_{P}(1)$. In addition, Assumptions \textbf{(A1)}%
-\textbf{(A5)} are enough to ensure that $\sup_{{\mathbf{\phi}}^{i}\in{{\Phi}%
}}\Vert\widehat{{\mathbf{\beta}}}(\mathbf{z(\phi}^{i}))-{\mathbf{b}%
}({\mathbf{\phi}}^{i})\Vert=o_{P}(1).$ The uniform convergence of
$\widehat{{\mathbf{\beta}}}(\mathbf{z(\phi}^{i}))$ to ${\mathbf{b}%
}({\mathbf{\phi}}^{i})$ is crucial as it ensures that the simulated paths
$\mathbf{z(\phi}^{i})$, and the subsequent $\widehat{\mathbf{\beta}%
}(\mathbf{z(\phi}^{i}))$,\ are well-behaved over\textbf{\ }${{\Phi}}$.
Assumptions (\textbf{I1})-(\textbf{I3}) ensure the required concentration of
the ABC posterior on sets containing the truth, ${\mathbf{\phi}}_{0}$. {In
particular, Assumption (\textbf{I1}) ensures that the prior used within ABC
places sufficient mass on the truth, and (some version of) this assumption is
standard\ in the analysis of Bayesian consistency.} {Assumption (\textbf{I3})
}is a type of deviation control for the estimated auxiliary parameters, and
allows us precise control over certain remainder terms in the posterior decomposition.

The following theorem formally establishes Bayesian consistency of the ABC
posterior in the SSM setting that is our interest herein.

\begin{theorem}
\label{Thm1} For all $\delta>0$, if Assumptions \textbf{(A)} and \textbf{(I)}
are satisfied, then, so long as $\varepsilon_{T}=o(1)$ is such that
$\varepsilon_{T}^{d_{\beta}+\kappa}v_{T}^{\kappa}\rightarrow\infty$, and
$\mathbf{\Omega}$ is positive definite,
\[
\int_{A_{\delta}({\mathbf{\phi}}_{0})}p_{\varepsilon}({\mathbf{\phi}%
}|{\mathbf{\eta}}({\mathbf{y}}))d{\mathbf{\phi}}=o_{P}(1),{\text{ for }%
}\;{\mathbf{\eta}}({\mathbf{y}})=\widehat{{\mathbf{\beta}}}({\mathbf{y}%
}),{\text{ as }}T\rightarrow\infty,
\]
where $A_{\delta}({\mathbf{\phi}}_{0}):=\{{\mathbf{\phi}}\in{{\Phi}}:d\left\{
{\mathbf{\phi}},{\mathbf{\phi}}_{0}\right\}  >\delta\}$.\footnote{The distance
in \eqref {dist_mle} essentially mimics the Wald criterion used in the
indirect inference technique. Similar to {the latter}, in our Bayesian
analyses, in which (\ref{dist_mle}) is used to produce ABC draws,
${\mathbf{\Omega}}$ can also be defined as the sandwich form of a
variance-covariance estimator (Gleim and Pigorsch, 2013, and Drovandi
\textit{et al.}, 2015), or as the inverse of the (estimated)
variance-covariance matrix for$\ \mathbf{\beta}$, evaluated at
$\widehat{{\mathbf{\beta}}}({\mathbf{y)}}$ (Drovandi\textit{\ et al.,} 2011).
In these cases it is more useful to denote the weighting matrix by
$\widehat{\boldsymbol{\Omega}}({\mathbf{y}},{\mathbf{\widehat{\beta
}({\mathbf{y}})}}) $ and Bayesian consistency then requires, in addition to
Assumptions\textbf{\ (A) }and\textbf{\ (I)},\textbf{\ }$\Vert
\widehat{\boldsymbol{\Omega}}({\mathbf{y}},\widehat{{\mathbf{\beta}}%
}({\mathbf{y)}})-\boldsymbol{\Omega}_{\infty}({\mathbf{\beta}}_{0})\Vert
_{\ast}\xrightarrow{P}0,$ for some positive definite $\boldsymbol{\Omega
}_{\infty}({\mathbf{\beta}}_{0})$, where\textbf{\ }$\Vert\mathbf{W}\Vert
_{\ast}=\sqrt{\text{Trace}(\mathbf{W}^{\prime}\mathbf{W})}$ for $\mathbf{W}$
an arbitrary $n\times m$ matrix.}
\end{theorem}

\noindent\textbf{Remark 2: }We have presented the conditions for consistency,
and proven Theorem 1, for the specific setting which is the focus here, namely
where both the true and auxiliary models are SSMs. The sufficient conditions
to ensure ${\mathbf{\eta}}({\mathbf{y}})=\widehat{{\mathbf{\beta}}%
}({\mathbf{y}})\xrightarrow{P}\mathbf{\beta}_{0},$ and, uniformly in
$\mathbf{\phi}^{i}$, $\mathbf{\eta}(\mathbf{z(\phi}^{i}%
))=\widehat{{\mathbf{\beta}}}(\mathbf{z(\phi}^{i}){\mathbf{)}}%
\xrightarrow{P}{\mathbf{b}}({\mathbf{\phi}}^{i}{\mathbf{)}}$ - \textbf{(A1)}
to \textbf{(A5)} - are based on the\textbf{\ }conditions invoked by Douc and
Moulines (2012) to establish consistency of the MLE in misspecified SSMs.
Whilst these authors use simple examples to illustrate their theory, in our
ABC setting, in which the true data generating process is, by the very nature
of the exercise, a challenging one, {analytical} verification of these
conditions is typically not possible. Similarly, it would appear to be
infeasible to verify \textbf{(I3) }analytically under the remaining maintained
assumptions in the {usual} case in which\ $\widehat{{\mathbf{\beta}}%
}(\mathbf{z}(\mathbf{\phi}))$ is unavailable in closed form. Moreover, and in
common to all simulation-based inference procedures, analytical verification
of the identification condition in \textbf{(I2)} is infeasible as a general
rule, and, hence, remains an open problem. Nevertheless, in the
{supplementary} {material} we demonstrate numerically the
Bayesian consistency of the auxiliary likelihood-based ABC method in all three
classes of SSM that we investigate therein.\footnote{We refer the interested
reader to Lomdardi and Calzolari (2009) for discussion of the difficulties of
verifying sufficient conditions for the asymptotic properties of indirect
inference methods in general, as well as in more specific, contexts.}

\subsection{Computationally Efficient ABC}

\subsubsection{Score-based ABC implementation}

With large computational gains, $\mathbf{\eta}(\mathbf{.})$ in (\ref{distance}%
) can be defined using the score of the auxiliary model. (See Gouri\'{e}roux
and Monfort, 1995, and Gallant and Tauchen, 1996 for the comparable point
first being made in the indirect inference context.)\textbf{\ }That is, the
score vector associated with the approximating model, when evaluated at the
simulated data, and with $\widehat{\mathbf{\beta}}(\mathbf{y)}$ substituted
for $\mathbf{\beta}$, will be closer to zero the `closer' is the simulated
data to the observed data. Hence, the distance in (\ref{distance}) can be
replaced by\textbf{\ }%
\begin{equation}
d\{\mathbf{\eta}(\mathbf{y}),\mathbf{\eta}(\mathbf{z(\phi}^{i}))\}=\sqrt
{\left[  \mathbf{S}(\mathbf{z(\phi}^{i});\widehat{\mathbf{\beta}}%
(\mathbf{y)})\right]  ^{\prime}\mathbf{\Sigma}\left[  \mathbf{S}%
(\mathbf{z(\phi}^{i});\widehat{\mathbf{\beta}}(\mathbf{y)})\right]  },
\label{distscore}%
\end{equation}
where%
\begin{equation}
\mathbf{S}(\mathbf{z(\phi}^{i});\mathbf{\beta})=T^{-1}\frac{\partial
L_{a}(\mathbf{z(\phi}^{i});\mathbf{\beta})}{\partial\mathbf{\beta}} 
\label{score}%
\end{equation}
is\textbf{\ }the (average) score of the auxiliary likelihood, $\mathbf{S}%
(\mathbf{y};\widehat{\mathbf{\beta}}(\mathbf{y)})=0,$ and $\mathbf{\Sigma}%
$\ denotes a positive definite weighting matrix which, if an estimated
quantity, satisfies comparable conditions to those specified in {Footnote 3}
for $\widehat{\boldsymbol{\Omega}}(.){\mathbf{.}}$ Implementation of ABC via
(\ref{distscore}) is faster (by orders of magnitude) than the approach based
upon $\mathbf{\eta}(.)=\widehat{\mathbf{\beta}}(.)$, due to the fact that
maximization of the auxiliary likelihood is required only once, in order to
produce $\widehat{\mathbf{\beta}}(.)$ from the observed data $\mathbf{y.}$ All
other calculations involve simply the evaluation\emph{\ }of $\mathbf{S}%
(.;\widehat{\mathbf{\beta}}(\mathbf{y)})$ at the simulated data, with a
numerical differentiation technique invoked to specify $\mathbf{S}%
(.;\widehat{\mathbf{\beta}}(\mathbf{y)}),$ when not known in closed form.

Similar to Theorem 1, we can demonstrate consistency of the ABC posterior
based on the auxiliary score. This result requires similar assumptions to
Theorem 1, but requires the following specific {variants} of Assumptions
\textbf{(A5)}, and \textbf{(I2)} and \textbf{(I3)}. \bigskip

\noindent\textbf{(A5$^{\prime}$)} (i) $\mathbf{S}_{\infty}(\mathbf{\phi
}\mathbf{;\beta}):=$ $\text{plim}_{T\rightarrow\infty}\mathbf{S}%
(\mathbf{z(\phi}_{{}});\mathbf{\beta})$ exists for all $\mathbf{\phi}$, where
$\mathbf{S}_{\infty}(\mathbf{\phi}^{i}\mathbf{;\beta}):=(\partial
/\partial\mathbf{\beta}^{\prime})L_{\infty}(\mathbf{\phi}^{i}\mathbf{,\beta}%
)$; (ii) $\mathbf{S}_{\infty}(\mathbf{\phi}\mathbf{;\beta}_{0})=0\text{ if and
only if }\mathbf{\phi=\phi}_{0}$. \bigskip

\noindent\textbf{(I2$^{\prime}$)} Let $\mathbf{\zeta}=(\mathbf{\phi}^{\prime
},\mathbf{\beta}^{\prime})^{\prime}$. For any $\mathbf{\zeta}\in\Phi
\times\mathcal{B}$, there exist some $C_{n}=O_{P}(1)$ such that \newline%
$\Vert\mathbf{S}(\mathbf{z(\phi}_{1}\mathbf{);\beta}_{1})-\mathbf{S}%
(\mathbf{z(\phi}_{2}\mathbf{);\beta}_{2})\Vert\leq C_{n}\Vert\mathbf{\zeta
}_{1}\mathbf{-\zeta}_{2}\Vert$.
%\begin{equation*}
%P_{z}^{\phi }\left[ \left\|
%S_{a}(\mathbf{z(\phi)};\mathbf{\beta })-S_{\infty}(\theta;\beta)  \right\|>u\right] \leq C(\zeta)u^{-\kappa }v_{T}^{-\kappa },\text{ and }\int_{\Phi }C(\mathbf{\zeta }%
%)\Pi (d\mathbf{\phi })<\infty,
%\end{equation*}with $C(\cdot)$ continuous for all  $\beta\in\mathcal{B}$.
\bigskip

\noindent\textbf{(I3$^{\prime}$)} For any $\mathbf{\phi}\in\Phi$, there exist
constants $\kappa,C,u_{0}>0$ such that, for some sequence $v_{T}%
\rightarrow\infty$ and all $0<u<u_{0}v_{T}$,
\[
P_{z}^{\phi}\left[  \left\Vert \mathbf{S}(\mathbf{z(\phi)};\mathbf{\beta}%
_{0})-\mathbf{S}_{\infty}(\mathbf{\phi};\mathbf{\beta}_{0})\right\Vert
>u\right]  \leq C(\mathbf{\phi})u^{-\kappa}v_{T}^{-\kappa},\text{ and }%
\int_{\Phi}C(\mathbf{\phi})\Pi(d\mathbf{\phi})<\infty.
\]

\begin{theorem}
\label{Thm2} If Assumptions \textbf{(A1)-(A4)}, \textbf{(A5$^{\,\prime}$)} and
\textbf{(I1)}, \textbf{(I2$^{\,\prime}$)}, \textbf{(I3$^{\,\prime}$)} are
satisfied, then, so long as $\varepsilon_{T}=o(1)$ is such that $\varepsilon
_{T}^{d_{\beta}+\kappa}v_{T}^{\kappa}\rightarrow\infty$, and
$\boldsymbol{\Sigma}$ is positive definite, Theorem \ref{Thm1} is satisfied
with $d\{\eta(\mathbf{y}),\eta(\mathbf{z}(\phi))\}=\sqrt{[\mathbf{S}%
(\mathbf{\phi;\widehat{\beta}(y)})]^{\prime}\boldsymbol{\Sigma}[\mathbf{S}%
(\mathbf{\phi;\widehat{\beta}(y)})]}.$
\end{theorem}

{In the supplementary material we informally demonstrate that, in addition to
the shared property of Bayesian consistency, for }$T\rightarrow\infty${\ and
}$\varepsilon_{T}\rightarrow0${, the score and MLE-based ABC selection
criteria will yield equivalent draws of }$\mathbf{\phi}${\ and, hence,
equivalent estimates of }$p(\mathbf{\phi}|\mathbf{y})${. As a consequence of
this (asymptotic) validity of the score-based method, in Section 5 we focus
entirely on this computationally efficient form of implementing auxiliary
likelihood-based ABC.}

\subsubsection{Dimension reduction via an integrated likelihood technique
\label{dimen-red}}

As highlighted by Blum (2010) (amongst others) the accuracy with which ABC
draws estimate the so-called partial posterior,\textbf{\ }$p(\mathbf{\phi
|\eta}(\mathbf{y}))$, for any given tolerance $\varepsilon$\ and number of
simulation draws $N$, will be less, the larger the dimension of $\mathbf{\eta
}(\mathbf{y})$. This `curse of dimensionality' obtains even when the parameter
$\mathbf{\phi}$ is a scalar, and relates solely to the dimension of
$\mathbf{\eta}(\mathbf{y}).$ As elaborated on further by Nott \textit{et al.}
(2014), this problem is exacerbated as the dimension of $\mathbf{\phi}$ itself
increases, firstly because an increase in the dimension of $\mathbf{\phi}$
brings with it a concurrent need for an increase in the dimension of
$\mathbf{\eta}(\mathbf{y})$ and, secondly, because the need to estimate a
multi-dimensional density (for $\mathbf{\phi}$) brings with it its own
problems related to dimension.\footnote{See Blum \textit{et al.} (2013) for
further elaboration on the dimensionality issue in ABC and a review of current
approaches for dealing with the problem. See also Frazier \textit{et al}.
(2018) for some {additional theoretical insights} into the issue.} This type
of inaccuracy is, of course, distinct from the inaccuracy that results from
the use of summary statistics that are not sufficient for\textbf{\ }%
$\mathbf{\phi}$.

We discuss here a dimension reduction technique that is particularly apt when
there is a natural link between the elements of the true and auxiliary
parameter vectors, and the dimensions of the two vectors ({assumed to be
greater than one}) are equivalent. In brief:\textbf{\ }let $\mathbf{\beta
}_{-j}=(\beta_{1},...,\beta_{j-1},\beta_{j+1},...,\beta_{d_{\phi}})^{\prime}$
be the $(d_{\phi}-1)$-dimensional parameter vector of auxiliary parameters,
and $\mathcal{B}_{-j}\subset\mathbb{R}^{(d_{\phi}-1)}$ be\textbf{\ }the
parameter space associated with $\mathbf{\beta}_{-j}$.\textbf{ }Letting
$w(\beta_{-j}|\beta_{j})$ denote a \textquotedblleft weight
function\textquotedblright\ for the auxiliary parameters $\beta_{-j}$, we can
define the integrated auxiliary likelihood $L_{a}^{I}(y;\beta_{j})$ as
\begin{equation}
L_{a}^{I}(\mathbf{y};\beta_{j})=\int_{\mathcal{B}_{-j}}L_{a}(\mathbf{y}%
;\mathbf{\beta})w(\mathbf{\beta}_{-j}|\beta_{j})d\mathbf{\beta}_{-j}.
\label{int_like}%
\end{equation}
For the given auxiliary model, $L_{a}^{I}(\mathbf{y};\beta_{j})$ can be used
to obtain a\textbf{\ }convenient \textit{scalar} summary statistic for use in
estimating the marginal posterior $p(\phi_{j}|\mathbf{y})$ via ABC,\textbf{\ }%
using (for example) the integrated score,\textbf{\ }%
\[
\mathbf{S}^{I}(\mathbf{z}(\mathbf{\phi});\widehat{\beta}_{j})=\frac
{\partial\log\left(  L_{a}^{I}(\mathbf{z}(\mathbf{\phi});\beta_{j})\right)
}{\partial\beta_{j}}|_{\beta_{j}=\widehat{\beta}_{j}},
\]
evaluated at\textbf{\ }$\widehat{\beta}_{j}=\arg\max_{\beta_{j}}L_{a}%
^{I}(\mathbf{y};\beta_{j}),$ where\textbf{\ }$\phi_{j}$ represents the\textbf{
}parameter in the data generating process that most closely matches the role
played by $\beta_{j}$ in the auxiliary model.{\footnote{As this is an
auxiliary likelihood, and thus does not need to have a strict interpretation
as an actual likelihood, the weight function can be chosen to ensure that the
integral in \eqref{int_like} can be calculated easily. In situations where the
space $\mathcal{B}_{-j}$ is relatively small, one can often take $w(\beta
_{-j}|\beta_{j})$ to be unity, which is the approach taken in the numerical
exercise in Section 5.1.}} If the marginal posteriors only are of interest,
then all $d_{\phi}$ marginals can be estimated in this way, with $d_{\phi}$
applications of $(d_{\phi}-1)$-dimensional integration required at each step
within ABC to produce the relevant score statistics. If the\textit{\ joint}
posterior of $\mathbf{\phi}$ were of interest, the sort of techniques
advocated by Nott \textit{et al.} (2014), amongst others, could be used to
yield joint inference from the estimated marginal posteriors.\footnote{In the
case where $\phi$ itself is a scalar and the number of parameters in an
auxiliary model is greater than one, application of this technique would
require a decision to be made as to which particular auxiliary parameter was
the most informative for $\phi$. The marginal score for that element of
$\mathbf{\beta}$ would used as the single summary statistic in the ABC
algorithm for selecting draws of $\phi.$}

{In Section \ref{numerical results} we demonstrate the use of this method in
the case where the auxiliary model is a discretized version of the true{
continuous-time model} and, thus, there is a natural (and one-to-one) link
between the two sets of parameters as a consequence. This is a very obvious
case in which this particular dimension reduction technique is suitable. The
small number of (true and auxiliary) parameters also means that the
deterministic numerical integration that is required at each iteration of ABC
in order to evaluate (\ref{int_like}) and, hence, the integrated score for use
in selecting each }$\phi_{j}${, is not computationally prohibitive. We note
that for a well-chosen auxiliary model, there is likely to be a
\textit{qualitative} link between the auxiliary and structural parameters
(e.g. location, scale, tail behavior, persistence) that can be exploited to
decide which univariate auxiliary score to use as the selection statistic for
any given\ }$\phi_{j}$.\textbf{ }However, we do not pursue this avenue in any
general sense in the paper.

\section{Auxiliary likelihood-based ABC: A case study of financial volatility
models \label{model_vol}}

\subsection{Overview}

In response to {the} now well-established empirical {characteristics of asset
return volatility (see, }Bollerslev \textit{et al.}, 1992, for a comprehensive
review) many alternative time-varying volatility models have been proposed,
with continuous-time stochastic volatility (SV) models - often augmented by
random jump processes - being particularly prominent of late. This focus on
the latter form of models is due, in part, to the availability of (semi-)
closed-form option prices, with variants of the {SV model of Heston} (1993)
becoming the workhorse of the empirical option pricing literature. Given the
challenging nature of the (non-central chi-squared) transitions in this model,
Bayesian analyses of it have typically proceeded by invoking{ Euler
discretizations} for both the measurement and state processes and applying
MCMC- or SMC-based techniques to that discretized model (e.g., Eraker, 2004;
Johannes\textbf{\ }\textit{et al}., 2009). It has also featured in the
indirect inference and efficient method of moments literatures, as a very
consequence of the difficulty of evaluating the exact likelihood (e.g.
Andersen \textit{et al}., 2002; Gallant and Tauchen, 2010). It is of interest,
therefore, to {explore} the proposed ABC method in the context of this form of
model, and this is the focus of Section \ref{SQ}.

In Sections \ref{alpha2} and \ref{alpha1} we then pursue two alternative
volatility models in which the distinctly non-Gaussian features of the
innovations to conditional returns are captured via the use of $\alpha$-stable
processes (Carr and Wu, 2003; Lombradi and Calzolari, 2009; Peters \textit{et
al}., 2012). With the $\alpha$-stable process not admitting a closed-form
representation for the density function,\textbf{\ }models in which it appears
present challenges for exact inference and are thus a prime candidate for
analysis via ABC, in particular given that such processes \textit{can} be
simulated via the algorithm proposed in Chambers \textit{et al.} (1976).

To facilitate the link between the general theoretical material presented thus
far and the specific examples to follow, we use the notation $\mathbf{\phi}$
({similarly, }$\mathbf{\beta}$) to denote the vector of parameters
characterizing the true (auxiliary) model in each case, despite the
interpretation of the parameters obviously differing from case to case. We
also use $y_{t}$ to denote {the observation} in each example, $x_{t}$ to
denote the latent state and $w_{t}$ and $v_{t}$ to denote the measurement and
state error {respectively}, as is consistent with the notation defined in
(\ref{d1}) and (\ref{d2}).

\subsection{Square root stochastic volatility\label{SQ}}

In this section we begin by assuming an observed (de-meaned) logarithmic
return, $r_{t}$, with the square root model for\textbf{\ }the
variance\textbf{\ }$x_{t}$,
\begin{align}
r_{t}  &  =x{_{t}^{1/2}}\eta_{t},\label{ret_1}\\
dx_{t}  &  =(\phi_{1}-\phi_{2}x_{t})dt+\phi_{3}\sqrt{x_{t}}v_{t},
\label{Heston}%
\end{align}
where $v_{t}=dW_{t}$ is a Brownian increment, and $\eta_{t}$ is defined as an
$i.i.d.$ random variable with zero mean and variance 1. We observe a discrete
sequence of returns, and our goal is to conduct Bayesian inference on the
parameters governing the dynamics of volatility. We restrict the structural
parameters as $2\phi_{1}\geq\phi_{3}^{2}$ to ensure positive volatility, and
for some\textbf{\ }$M,\varphi,$\textbf{\ }we impose\textbf{\ }$M\geq\phi
_{3},\phi_{1},\phi_{2}\geq\varphi>0$. With these restrictions,\thinspace
$x_{t}$ is mean reverting and as $t\rightarrow\infty$, $x_{t}$ approaches a
steady state gamma distribution, with $\mathbb{E}[x_{t}]=\phi_{1}/\phi_{2}$
and $var(x_{t})=\phi_{3}^{2}\phi_{1}/2\phi_{2}^{2}.$ The conditional
distribution function for $x_{t}$ is {non-central chi-squared}, $\chi
^{2}(2cx_{t};2q+2,2u)$, with $2q+2$ degrees of freedom and non-centrality
parameter $2u$. The transition density for $x_{t}$, conditional on $x_{t-1}$,
is thus
\begin{equation}
p(x_{t}|x_{t-1},\mathbf{\phi})=c\exp(-u-v)\left(  \frac{v}{u}\right)
^{q/2}I_{q}(2(uv)^{1/2}), \label{Bessel}%
\end{equation}
where $c=2\phi_{2}/\phi_{3}^{2}(1-\exp(-\phi_{2}))$, $u=cx_{t-1}\exp(-\phi
_{2})$, $v=cx_{t}$, $q=\frac{2\phi_{1}}{\phi_{3}^{2}}-1$, and $I_{q}(.)$ is
the modified Bessel function of the first kind of order $q.$

With both the conditional density in (\ref{meas_gen}) and the transition
density in (\ref{state_gen}) being available for this model, likelihood-based
inference is, in principle, feasible. {However}, we view the application of
ABC in this setting as an attractive option to explore, in particular given
the ability to \textit{simulate} the process via\textbf{\ }its representation
as a composition of ({central) chi-squared and Poisson distributions. {In
Section \ref{sq_numerics}, in order to produce an exact comparator for the ABC
posterior estimate for this model, we exploit the availability of the
transition densities in (\ref{Bessel}) and }apply the non-linear filter of Ng
\textit{et al. }(2013) to evaluate the likelihood, {thereafter normalizing}
the exact posterior using deterministic numerical integration techniques. {We}
do not propose the latter as a computationally attractive (or readily
generalizable) competitor to the ABC approach, simply using it in a one-off
exercise for the purpose of evaluation. }

For convenience,\textbf{\ }we take squares and logarithms of the measurement
equation to define
\begin{align}
y_{t} &  =\ln(r_{t}^{2})-\omega=\ln(x_{t})+w_{t},\label{sq_1}\\
dx_{t} &  =(\phi_{1}-\phi_{2}x_{t})dt+\phi_{3}\sqrt{x_{t}}v_{t},\label{sq_2}%
\end{align}
where $w_{t}=\ln(\eta_{t}%
^{2})-\omega,$ with $\omega=\mathbb{E}[\ln(\eta_{t}^{2})]$. To implement an
auxiliary likelihood-based ABC algorithm, we {initially }adopt a Gaussian
approximation for $w_{t}$ in (\ref{sq_1}) and an Euler discretization for
(\ref{sq_2}), yielding the approximating model,
\begin{align}
y_{t} &  =\ln(x_{t})+\epsilon_{t},\label{sq_approx}\\
x_{t} &  =\beta_{1}+\beta_{2}x_{t-1}+\beta_{3}\sqrt{x_{t-1}}e_{t}%
,\label{sq_2_discrete}%
\end{align}
where $\epsilon_{t}\sim N(0,\sigma_{w}^{2})$, $e_{t}$ is a truncated Gaussian
variable with lower bound, $e_{t}>\frac{-\beta_{1}}{\beta_{3}},$ and we define
the auxiliary parameters as $\mathbf{\beta}=(\beta_{1},{\beta}_{2},{\beta}%
_{3})^{\prime}$. Similar parameter restrictions to those imposed on the
structural parameters $\mathbf{\phi}$\ are required of the elements of
$\mathbf{\beta}$: $M\geq\beta_{1},$ $\beta_{3}\geq\varphi>0$, $\varphi
\leq\beta_{2}\leq1-\varphi$, and $2\beta_{1}\geq\beta_{3}^{2}$. The equations
in (\ref{sq_approx}) and (\ref{sq_2_discrete}) play the role of (\ref{miss1})
and (\ref{miss2}) respectively.

The non-linearities that\textbf{\ }characterize both (\ref{sq_approx}) and
(\ref{sq_2_discrete}) imply that an analytical evaluation of the auxiliary
likelihood via the Kalman filter (KF) is not feasible. Therefore, we use the
augmented unscented KF (AUKF) (see Julier \textit{et al.}, 1995)\textbf{ }as
an computationally efficient means of evaluating the $L_{a}(\mathbf{y}%
;\mathbf{\beta})$ and,\textbf{\ }hence, of producing the auxiliary score as
the matching statistic within ABC. The precise form of the\ auxiliary
likelihood function thus depends on both\textbf{\ }the first-order Euler
discretization of the continuous-time state process \textit{and} the
particular specifications used to implement the AUKF. General pseudo code for
the AUKF is given in algorithmic form in the supplementary material, along
with certain detailed implementation instructions. For comparison we also
experiment with various alternative auxiliary models from the GARCH family.
Further details of these models are provided in Section \ref{data}.

\subsection{Conditionally $\alpha$-stable returns with stochastic volatility
\label{alpha2}}

Let $\{X_{t}^{\alpha,\gamma}:\;t\in\mathbb{R}_{+}\}$ be an $\alpha$-stable
L\'{e}vy process with location $\mu=0$, scale $\sigma=1$, tail index
$\alpha\in(1,2)$, and skewness parameter $\gamma\in\lbrack-1,1]$. Then $X_{t}$
has independent and stationary increments $dX_{t}^{\alpha,\gamma}$ such that
$dX_{t}^{\alpha,\gamma}\sim\mathcal{S}(\alpha,\gamma,0,dt^{1/\alpha})$ and
exhibits differing degrees of leptokurtosis and skewness depending on the
values of $\alpha$ and $\gamma$. The process is also self-similar in that the
distribution of an $\alpha$-stable variable defined over any horizon has the
same shape upon scaling. Critically however, the density function has no
closed-form representation. (See Samorodnitsky and Taqqu, 1994, Chapter 7.)

Recently, several authors have used $\alpha$-stable L\'{e}vy motion to model
financial data. Notably, Carr and Wu (2003) model logarithmic returns on the
S\&P500 price index as\textbf{\ }$\alpha$-stable, with a view to capturing the
lack of `flattening' of the implied volatility smile as option maturity
increases. In brief, the infinite variance (for the log return) implied by
this model violates the conditions for a Gaussian central limit theorem and,
hence, fits with the phenomenon of a smile that persists. At the same time,
however, with the lower bound imposed for $\gamma$, the conditional
expectation of the index itself remains finite, thereby enabling meaningful
European option prices to be defined. Whilst the detailed derivations in their
paper pertain to the case in which volatility is constant, recognition of the
need to incorporate stochastic volatility prompts {those} authors to propose
(as a vehicle for future research) an extended model in which the Heston
(1993) model in (\ref{Heston}) is adopted for the variance, with closed-form
option pricing still being feasible as a consequence.

Most importantly, with the focus in Carr and Wu (2003) being on the estimation
of risk neutral parameters via calibration of the model with market option
prices, the lack of analytical form for the density of $X_{t}$ is not a
hindrance for inference. However, any attempt to conduct likelihood-based
inference (including exact Bayesian inference) using spot returns\textbf{
}\textit{would} encounter this hurdle, with the conditional density in
(\ref{meas_gen}) being unavailable; and that is where ABC provides a useful alternative.

With this empirical motivation in mind, we thus explore the application of ABC
to the model%
\begin{align}
y_{t}  &  =r_{t}=x{_{t}^{1/\phi_{4}}}w_{t},\label{ret_2}\\
\ln x_{t}  &  =\phi_{1}+\phi_{2}\ln x_{t-1}+\phi_{3}v_{t}, \label{ar1}%
\end{align}
where $w_{t}\sim i.i.d.$ $\mathcal{S}(\phi_{4},-1,0,dt=1)$, $v_{t}\ $is an
$i.i.d.$ random variable (independent of $w_{t}$) with zero mean and variance
1, and to be consistent with our general notation, we denote the tail index
$\alpha$ by $\phi_{4}.$ {To ensure stationarity in the $x_{t}$ process, we
impose $|\phi_{2}|<1$, and $\phi_{3}$ is }required to be strictly positive.{
As in Carr and Wu (2003), we restrict $\phi_{4}\in(1,2]$ to ensure that
$y_{t}$ has support over the whole real line.} Once again we assume discretely
observed returns and, for the sake of illustration, work with a discrete-time
autoregressive model for the logarithm of the variance, as given in
(\ref{ar1}). In particular, this allows us to illustrate ABC using the
following simple auxiliary model based on a first-order generalized
autoregressive conditional heteroscedastic (GARCH(1,1)) model for the latent
standard deviation,
\begin{align}
y_{t}  &  =r_{t}=x{_{t}}\epsilon_{t},\label{miss1_2}\\
x_{t}  &  =\beta_{1}+\beta_{2}x_{t-1}\left\vert \epsilon_{t-1}\right\vert
+\beta_{3}x_{t-1}, \label{miss2_2}%
\end{align}
where $\epsilon_{t}\sim i.i.d.$ $St(0,1,\beta_{4}).$ That is, the measurement
error in the auxiliary model is a standardized Student $t$ random variable
with degrees of freedom parameter $\beta_{4}$. (See also Lombardi and
Calzolari, 2009, and Garcia \textit{et al., }2011, for the application of
indirect inference to similar model scenarios.) The ARCH component of
(\ref{miss2_2}) is parameterized using absolute deviations (instead of
squares) to mitigate numerical instabilities that can arise from extreme
realizations of the $\alpha$-stable distribution. {As with standard GARCH
models, we impose positivity of $\beta_{1},\beta_{2}$ and $\beta_{3}$ to
ensure positivity of $x_{t}$, and ensure stationarity by restricting
$\beta_{2}+\beta_{3}<1$.} {Note that, by defining }$e_{t}=\left\vert
\epsilon_{t-1}\right\vert ${, and with an appropriate transformation}, the
model in (\ref{miss1_2}) and (\ref{miss2_2}) can be placed in the state space
form given in (\ref{miss1}) and (\ref{miss2}); but with the auxiliary
likelihood function available in closed form in this case, the application of
ABC is particularly straightforward and does not require
filtering.\footnote{Use of the square root volatility model in (\ref{ar1})
would also of course be feasible, but the heteroscedastic nature of the
variance model would demand an auxiliary model that reflected that feature,
{along the lines of (\ref{sq_2})}, and hence, entail the use of filtering to
evaluate the auxiliary likelihood.}

\subsection{Stochastic volatility with $\alpha$-stable errors \label{alpha1}}

An alternative approach to modelling the stylized features of financial
returns is to consider a stochastic volatility model for returns in which an
$\alpha$-stable process drives the innovations to (log) volatility itself; see
Lombardi and Calzolari (2009) once again. To that end, in this section we
define the following model for the return,
\begin{align}
r_{t}  &  =x{_{t}^{1/2}}w_{t},\label{ret_3}\\
\ln x_{t}  &  =\phi_{1}+\phi_{2}\ln x_{t-1}+\phi_{3}v_{t}, \label{ar_3}%
\end{align}
where $v_{t}\sim i.i.d.$ $\mathcal{S}(\phi_{4},-1,0,dt=1)$, and $w_{t}\ $is an
$i.i.d.$ random variable (independent of $v_{t}$) with zero mean and variance.
Again, stationarity of $x_{t}$ requires $|\phi_{2}|<1$, and positivity of
$x_{t}$ requires $\phi_{3}>0$. Similar to the previous $\alpha$-stable
example, we restrict $\phi_{4}\in(1,2]$. With this particular specification it
is the transition density in (\ref{state_gen}) that is unavailable, rendering
exact likelihood-based inference infeasible. In the spirit of Lombardi and
Calzolari we base ABC on a (conventional) GARCH(1,1) auxiliary model for the
latent variance:\textbf{\ }%
\begin{align}
y_{t}  &  =r_{t}=x{_{t}^{1/2}}\epsilon_{t},\label{aux_ret_3}\\
x_{t}  &  =\beta_{1}+\beta_{2}x_{t-1}\epsilon_{t-1}^{2}+\beta_{3}x_{t-1},
\label{aux_garch_3}%
\end{align}
in which case {the restrictions on $\beta_{1},\beta_{2}$ and $\beta_{3}$, as
well as }the computational burden of the ABC method, {are} comparable to that
in Section \ref{alpha2}.

\section{Numerical assessment of auxiliary likelihood-based ABC\label{assess}}

\subsection{Overview}

We undertake {two} numerical exercises in which the {performance} of the
auxiliary likelihood-based approach to ABC is documented. As noted earlier,
having established its asymptotic validity for this purpose, we use the
auxiliary score to define the set of summary statistics, eschewing the direct
use of the more computationally burdensome auxiliary MLE.\footnote{As further
motivation for this decision, we note that a small numerical exploration
undertaken using certain specifications underpinning the numerical exercises
{documented in Section \ref{sq_numerics}} produced computation times
for the auxiliary MLE (as summary) that were approximately 60 times
greater\textbf{ }than corresponding computational times for the auxiliary
score, simply due to the need to optimize the auxiliary likelihood function at
each iteration of ABC.} The {first
exercise}, in Section \ref{sq_numerics}, uses the model in (\ref{sq_1}) and
(\ref{sq_2}) as the example, with a Gaussian assumption adopted for the
conditional distribution of returns, with this model referred to as SV-SQ
hereafter. With the exact posterior accessible in this case ({via the
deterministic non-linear filtering technique} {of }Ng \textit{et al}.,
2013) we are able to document the {accuracy} of a range of different ABC
estimates of the exact marginal posterior densities, both with and without the
use of dimension reduction of some sort. Both accept/reject and MCMC-based ABC
algorithms are applied, as are algorithms based on {a vector of }statistics
that {does }{{not exactly correspond }to a (vector) }auxiliary score. Two
particle marginal Metropolis Hastings (PMMH) estimates of the exact marginals
are also produced, for the purpose of comparison. {A second }numerical
exercise, which is presented in the supplementary appendix due to space
restrictions, explores the large sample behaviour of certain ABC posterior
estimates, for all three classes of stochastic volatility model. In
particular, {in this additional exercise}, we numerically verify the
conclusions\textit{ }of Theorem \ref{Thm2} by {demonstrating that the
auxiliary-score based }ABC estimates concentrate on the true parameter values
as the sample size increases.

\subsection{Finite sample accuracy of marginal posterior density estimation:
the SQ-SV model\label{sq_numerics}}

\subsubsection{Data generation and computational details\label{data}}

For the purpose of this illustration we{ simulate} an `empirical' sample of
size $T$ from the model in\ (\ref{sq_1}) and (\ref{sq_2}), with the parameters
set to values that yield\textbf{\ }observations on both $r_{t}$ and $x_{t}$
that match the characteristics of (respectively) daily returns and daily
values of realized volatility (constructed from 5 minute returns) for the
S\&P500 stock index over the 2003-2004 period: namely, $\phi_{1}=0.004;$
$\phi_{2}=0.1;$ $\phi_{3}=0.062.$ Choosing this relatively calm period in the
stock market as a reference point obviates the need to augment the SV-SQ model
with random price jumps and/or a non-Gaussian conditional distribution, and
enables the known form of the SV-SQ transition densities to be used in
producing the exact comparator via the method of Ng \textit{et al.}
(2013).\textbf{\ }In brief, the algorithm of Ng \textit{et al. }represents the
recursive filtering and prediction distributions used to define the
exact\textbf{\ }likelihood function as the numerical solutions of integrals
defined over the support of $w_{t}$ in (\ref{sq_1}), with deterministic
integration used to evaluate the relevant integrals, and the exact transitions
in (\ref{sq_2}) used in the specification of the filtering and up-dating
steps. Whilst lacking the general applicability of an ABC-based approach, this
deterministic filtering method is ideal for the particular model used in this
illustration, and can be viewed as producing a very accurate estimate of the
exact density, without any of the simulation error that would be associated
with a PMMH-based comparator, for instance. We refer the reader to Ng
\textit{et al.} for more details of the technique. The likelihood function,
evaluated via this method, is then multiplied by a uniform prior that imposes
the restrictions: $0.5<\phi_{2}<1$; $0.002<\phi_{1}<0.025,$ $0.005<\phi
_{3}<0.89$ and $2\phi_{1}\geq\phi_{3}^{2}$. The three marginal posteriors are
produced via deterministic numerical integration (over the parameter space),
with a very fine grid on $\mathbf{\phi}$ being used to ensure accuracy. These
marginals are used as the benchmark for assessing the accuracy of all
competing density estimates.

The auxiliary likelihood function (and score) of the approximating model
defined by (\ref{sq_approx}) and (\ref{sq_2_discrete}) is evaluated\ using the
AUKF, in the manner described in Section \ref{SQ} (and in the supplementary
material). We also explore the performance of the auxiliary score method using
a range of GARCH-type auxiliary models with closed-form likelihood functions,
details of which are provided in Table 1. Hence, five alternative auxiliary
models are explored {in total}. In all cases, the ABC score-based method uses
the distance measure in (\ref{distscore}).\textbf{\ }The weighting matrix
$\mathbf{\Sigma}$ is set equal to the Hessian-based estimate of the
variance-covariance matrix of the (joint)\textbf{\ }MLE of $\mathbf{\beta}$,
evaluated at the MLE computed from the observed data, $\widehat{\mathbf{\beta
}}(\mathbf{y).}$%

%TCIMACRO{\TeXButton{B}{\begin{table}[tbp] \centering}}%
%BeginExpansion
\begin{table}[tbp]\label{GARCH}\caption{
{\small Auxiliary models from the GARCH family. The abbreviations are defined
as follows: GARCH-N: GARCH with normal error, }$\epsilon_{t}${\small ;
GARCH-T: GARCH with Student }$t$ {\small error, }$\epsilon_{t}${\small ;
TGARCH-N: Threshold GARCH with normal error, }$\epsilon_{t}${\small ;
TGARCH-T: Threshold GARCH with Student }$t$ {\small error, }$\epsilon_{t}$}
\begin{tabular}
[c]{lrcc}\hline\hline
\textbf{Abbreviation} & \multicolumn{3}{c}{\textbf{Auxiliary Model
Specification}}\\\hline\hline
GARCH-N & $y_{t}=r_{t}=x_{t}^{1/2}\epsilon_{t};$ & \multicolumn{1}{l}{$x_{t}%
=\beta_{0}+\beta_{1}x_{t-1}\epsilon_{t}^{2}+\beta_{2}x_{t-1};$} &
\multicolumn{1}{l}{$\epsilon_{t}\sim N(0,1)$}\\
GARCH-T & $y_{t}=r_{t}=x_{t}^{1/2}\epsilon_{t}$; & \multicolumn{1}{l}{$x_{t}%
=\beta_{0}+\beta_{1}x_{t-1}\epsilon_{t}^{2}+\beta_{2}x_{t-1};$} &
\multicolumn{1}{l}{$\epsilon_{t}\sim t(\nu)$}\\
TARCH-N & $y_{t}=r_{t}=x_{t}^{1/2}\epsilon_{t};$ & \multicolumn{1}{l}{$x_{t}%
=\beta_{0}+\beta_{1}x_{t-1}\epsilon_{t}^{2}+\beta_{2}I_{(r_{t-1}<0)}%
x_{t-1}\epsilon_{t}^{2}+\beta_{3}x_{t-1};$} & \multicolumn{1}{l}{$\epsilon
_{t}\sim N(0,1)$}\\
TARCH-T & $y_{t}=r_{t}=x_{t}^{1/2}\epsilon_{t};$ & \multicolumn{1}{l}{$x_{t}%
=\beta_{0}+\beta_{1}x_{t-1}\epsilon_{t}^{2}+\beta_{2}I_{(r_{t-1}<0)}%
x_{t-1}\epsilon_{t}^{2}+\beta_{3}x_{t-1};$} & \multicolumn{1}{l}{$\epsilon
_{t}\sim t(\nu)$}\\\hline\hline
\end{tabular}
%TCIMACRO{\TeXButton{E}{\end{table}}}%
%BeginExpansion
\end{table}%
%EndExpansion

We compare the performance of the auxiliary score approaches with that of more
conventional ABC methods based on summary statistics that may be deemed to be
a sensible choice in this setting. For this purpose we propose a set of
summary statistics that are sufficient (under Gaussianity) for an observable
AR(1) process for the log of squared daily returns,\textbf{\ }$y_{t}=\ln
(r_{t}^{2})$, namely%
\begin{equation}
s_{1}=\sum_{t=2}^{T-1}y_{t},\text{ }s_{2}=\sum_{t=2}^{T-1}y_{t}^{2},\text{
}s_{3}=\sum_{t=2}^{T}y_{t}y_{t-1},\text{ }s_{4}=y_{1}+\text{ }y_{T},\text{
}s_{5}=y_{1}^{2}+y_{T}^{2}. \label{summaries}%
\end{equation}
For comparison we also compute the summaries from the raw (not transformed)
returns data.

Four different dimension reduction techniques are also\textbf{ }applied: 1)
the integrated likelihood technique described in Section \ref{dimen-red}; 2)
the linear regression adjustment method of Beaumont \textit{et al}. (2002); 3)
the neural network-based non-linear adjustment method of Blum and Francois
(2010); and 4) the semi-automatic procedure of Fearnhead and Prangle (FP)
(2012), based on polynomial basis functions up to the fourth order. Method 1)
is applied in the case where the scores are constructed from the AUKF-based
auxiliary likelihood function for (\ref{sq_approx}) and (\ref{sq_2_discrete}),
in which the dimensions of the true and auxiliary parameter sets are equal and
selection of each true parameter is based on the marginal score of the\textbf{
}\textit{corresponding}\textbf{ }parameter in the discretized approximation
model. Method 2) is applied to the scores of each of the five different
auxiliary models, and is also implemented jointly on the scores of all four
GARCH models. Method 3) is also applied to the combined scores of the four
GARCH models. Method 4) is applied to the combined GARCH-model scores, and,
separately, to the summaries in (\ref{summaries}) computed from both the
transformed and raw data.

In addition to the above ABC\ approaches, in which the accept/reject form of
Algorithm 1 is applied, we consider an implementation of ABC-MCMC (Marjoram
\textit{et al.}, 2003). ABC-MCMC\ replaces the discontinuous rejection step in
Algorithm 1 with a random walk MH step, in order to explore the posterior
support more efficiently. To keep the extent of the numerical results within
reason, we implement this alternative version of ABC only in the case of the
four auxiliary GARCH models.

The AUKF evaluation of the likelihood function (and joint score)
associated with (\ref{sq_approx}) and (\ref{sq_2_discrete}) is performed
using the GAUSS software, with the integration required to produce the
marginal score function implemented using a numerical integration subroutine
from \texttt{C}. All other computations are implemented in MATLAB and
\texttt{{R}}.\footnote{Computations related to estimation methods 4-7 in Table 3 are performed in MATLAB. {Computations related to the dimension reduction techniques applied to the GARCH and AUKF scores (generated by MATLAB and GAUSS respectively), are undertaken using the }\texttt{R} packages: \texttt{abc} (Csillery \textit{et al}., 2012) and \texttt{abctools}\textbf{ }(Nunes and Prangle, 2015); with packages \texttt{EasyABC} (Jabot \textit{et al.}, 2015) and \texttt{rugarch} (Ghalanos, 2018) used for the ABC-MCMC methods.} Each instance of ABC is based on $N=50,000$\ simulated draws from uniform
priors on $\phi_{1},$ $\phi_{2}$ and $\phi_{3}$ truncated as described above.
Draws are retained that lead to distances within the 0.5\% quantile of the
overall simulated distances, which are intrinsic to each procedure.

Finally, two filtering-based methods are used to produce an estimate of the
exact posterior via PMMH. The random walk MH algorithm is used, with the
likelihood in the MH ratio computed by 1) the bootstrap particle filter; and
2) the ABC filter as per Fig.1 of Jasra \textit{et al}. (2012). We denote the
posteriors produced by these methods as PMMH-BPF and PMMH-ABCF, respectively.
With the bootstrap filter requiring only simulation from the transitions in
(\ref{Bessel}), up to simulation error arising from both the use of a finite
number of particles and a finite number of (autocorrelated) MH draws, the
PMMH-BPF can be viewed as providing `exact' estimates of the marginal
posteriors. The PMMH-ABCF might also be described as such, but only
conditional on the tolerance used in the selection step being {sufficiently
close to zero}. However PMMH-ABCF (as applied here) certainly implements ABC
without data summarization and, hence, can be viewed as avoiding that
\textit{particular} disconnect from the exact posterior.\footnote{Discussion
of the asymptotic properties (including as the tolerance declines to zero) of
algorithms that employ an ABC filtering step can be found in Jasra (2015).}
Both filtering-based estimates are produced from 10,000 draws following 5,000
burn-in draws, {and with 3,000 particles used in each instance of likelihood
estimation. {The unknown parameters are drawn as a block, using truncated
normal proposals formed based on the previous draw, adhering to the relevant
model's parameter restrictions and prior boundaries.} The application of the
ABC filter uses a tolerance that ensures that only simulated draws that are
within 30\% of the actual observation at any time }$t$ are
retained{.\footnote{We computed the likelihood using the ABC filter over a
range of tolerance levels, and selected the tolerance level at which point the
likelihood estimate became stable.}} The filtering-based computations are all performed in
\texttt{{R}}, {with code provided in the supplementary material}.

\subsubsection{Numerical accuracy of marginal posterior estimates
\label{numerical results}}

To aid the reader, we begin by producing in Table 2 a key to the 18 methods we
use to estimate the exact (marginal) posterior for each parameter in the SV-SQ
model. In Table 3, we then summarize the accuracy of each approach by
reporting the root mean squared error (RMSE) for a given parameter, computed
as:%
\begin{equation}
RMSE=\sqrt{\frac{1}{G}\textstyle\sum\limits_{g=1}^{G}(\widehat{p}_{g}%
-p_{g})^{2}}, \label{rmse_1}%
\end{equation}
where $\widehat{p}_{g}$ is the ordinate of the relevant density estimate
(produced using kernel density methods) and $p_{g}$ the ordinate of the exact
posterior density, calculated using the deterministic filtering approach of Ng
\textit{et al}. (2013), at the $gth$ grid-point in the support of the
parameter. The RMSE associated with a given estimation method, for any
particular parameter, is reported as a ratio of that method's RMSE relative to
the method with the smallest RMSE for that given parameter. The ranking of all
18 methods, in terms of the average RMSE across all three parameters, is
reported in the final column of Table 3.%

%TCIMACRO{\TeXButton{B}{\begin{table}[tbp] \centering}}%
%BeginExpansion
\begin{table}[h] \label{TableKey}
%EndExpansion
\caption{
{\small Summary description of the 18 {posterior sampling} methods.
AR-ABC is accept/reject ABC (i.e. Algorithm 1). AUKF refers to the score of
the auxiliary model in (22) and (23), evaluated using the AUKF. IN-AUKF refers
to the integrated score for each parameter in the auxiliary model in (22) and
(23), evaluated using the AUKF. The FP procedure uses fourth-order polynomial
basis functions.}}
{\small
\begin{tabular}
[c]{ll}\hline\hline
\textbf{Abbreviation} & \textbf{Details of the posterior sampler}\\\hline
AUKF-AR & AR-ABC with the joint AUKF auxiliary score and no dimension
reduction\\
AUKF-AR-LL & AR-ABC with the joint AUKF auxiliary score and linear reg. adj.\\
AUKF-AR-IN & AR-ABC with the integrated AUKF auxiliary score\\
GARCH-N-AR-LL & AR-ABC with the joint GARCH-N score and linear reg. adj.\\
GARCH-T-AR-LL & AR-ABC with the joint GARCH-T score and linear reg. adj.\\
TARCH-N-AR-LL & AR-ABC with the joint TGARCH-N score and linear reg. adj.\\
TARCH-T-AR-LL & AR-ABC with the joint TGARCH-T score and linear reg. adj.\\
Pooled-GARCH-AR-LL & AR-ABC with the scores from all four GARCH models and
linear reg. adj.\\
Pooled-GARCH-AR-NN & AR-ABC with the scores from all four GARCH models and
non-linear reg. adj.\\
GARCH-N-MCMC & ABC-MCMC with the joint GARCH-N score and no dimension
reduction\\
GARCH-T-MCMC & ABC-MCMC with the joint GARCH-T score and no dimension
reduction\\
TARCH-N-MCMC & ABC-MCMC with the joint TGARCH-N score and no dimension
reduction\\
TARCH-T-MCMC & ABC-MCMC with the joint TGARCH-T score and no dimension
reduction\\
FP-ABC-All & FP with all 16 scores from the four GARCH auxiliary models\\
FP-ABC-RAW & FP with the statistics in (\ref{summaries}), computed from the
raw data ($y_{t}=r_{t})$\\
FP-ABC-TRANS & FP with the statistics in (\ref{summaries}), computed from the
transformed data ($y_{t}=\ln
(r_{t}^{2})$)\\
PMMH-ABC & PMMH with the ABC particle filter\\
PMMH-BPF & PMMH with the bootstrap particle filter\\\hline\hline
\end{tabular}
}

%TCIMACRO{\TeXButton{E}{\end{table}}}%
%BeginExpansion
\end{table}%

%TCIMACRO{\TeXButton{B}{\begin{table}[tbp] \centering}}%
%BeginExpansion
\begin{table}[h] \label{Table1}
%EndExpansion

\caption{{\small RMSE of estimated marginal posterior densities. Results are reported
as the ratio of a method's RMSE relative to the smallest RMSE
for that given parameter. The ratio in bold indicates the most accurate method
for each parameter. The final column ranks the methods according to their
average RMSE over the marginal posteriors estimates.}}

{\small
\begin{tabular}
[c]{l|ccccccc}\hline\hline
\textbf{Parameter:} &  & $\phi_{1}$ &  & $1-\phi_{2}$ & $\phi_{3}$ &  &
\\\hline
\textbf{Estimation method} &  & \multicolumn{4}{c}{\textbf{Relative RMSE}} &  & \textbf{Overall rank}\\\hline
1. AUKF-AR &  & 1.9704 &  & 2.5427 & 1.3998 &  & 6\\
2. AUKF-AR-LL &  & 1.8997 &  & 2.2830 & 1.3312 &  & 5\\
3. AUKF-AR-IN &  & \textbf{1.0000} &  & 2.2004 & 1.3307 &  & \textbf{1}\\
4. GARCH-N-AR-LL &  & 2.2452 &  & 3.4830 & 1.2481 &  & 8\\
5. GARCH-T-AR-LL &  & 2.0812 &  & 2.9282 & \textbf{1.0000} &  & 7\\
6. TARCH-N-AR-LL &  & 2.5662 &  & 3.8728 & 1.3666 &  & 12\\
7. TARCH-T-AR-LL &  & 2.3733 &  & 3.2770 & 1.0758 &  & 9\\
8. Pooled-GARCH-AR-LL &  & 2.5496 &  & 3.5139 & 1.6371 &  & 13\\
9. Pooled-GARCH-AR-NN &  & 2.7637 &  & 3.0090 & 1.5869 &  & 14\\
10. GARCH-N-ABC-MCMC &  & 3.4765 &  & 5.0755 & 2.0844 &  & 17\\
11. GARCH-T-ABC-MCMC &  & 3.3041 &  & 5.5014 & 2.2565 &  & 15\\
12. TARCH-N-ABC-MCMC &  & 3.7154 &  & 5.4578 & 2.2127 &  & 18\\
13. TARCH-T-ABC-MCMC &  & 3.4319 &  & 5.2219 & 1.9644 &  & 16\\
14. FP-ABC-All &  & 1.7822 &  & 4.0849 & 1.0797 &  & 4\\
15. FP-ABC-RAW &  & 2.4332 &  & 3.9050 & 1.2885 &  & 10\\
16. FP-ABC-TRANS &  & 1.0759 &  & \textbf{1.0000} & 1.9254 &  & 2\\
17. PMMH-ABC &  & 2.2587 &  & 3.5515 & 2.5745 &  & 11\\
18. PMMH-BPF &  & 1.3584 &  & 2.8303 & 1.6552 &  & 3\\\hline\hline
\end{tabular}
}

%TCIMACRO{\TeXButton{E}{\end{table}}}%
%BeginExpansion
\end{table}%
%EndExpansion

{Whilst we do not claim to have exhausted all possibilities in this exercise,
the broad sweep of techniques applied allows us to draw some conclusions
regarding the nature of ABC density estimation in the state space setting. We
summarize the key results as follows. (i) Application of the auxiliary score
technique to the discretized version of the {true continuous-time model},
allied with the integrated likelihood approach to dimension reduction, yields
the most accurate results overall, even in comparison with the `exact'
PMMH-BPF method. Indeed, the less computationally burdensome AUKF-AR and
AUKF-AR-LL methods are both in the top third, in terms of ranking, indicating
the importance of using - if possible - an auxiliary model that closely mimics
the }model assumed to have generated the observed data.{ (ii) The
semi-automatic dimension reduction technique of Fearnhead and Prangle (2012)
performs very well for two choices of base statistics,} and are even superior
{to the `exact' PMMH-BPF comparator in one case. The markedly better
performance of FP-ABC-TRANS relative to FP-ABC-RAW highlights the fact that
the accuracy of ABC depends \textit{both} on the use of informative summaries,
and the reduction of dimension, with the summaries in (\ref{summaries}) being
sufficient for an observable measure of volatility \textit{only} when the data
is transformed appropriately. In other words, the positive impact of dimension
reduction cannot offset a poorly chosen set of statistics. (iii) Despite the
previous remark, dimension reduction of some sort is seen to be important.
Indeed use of the more sophisticated ABC-MCMC algorithm does not compensate
for the lack of dimension reduction, }which is evidenced by the fact that the
ABC-MCMC methods occupy the lowest third of the rankings.{ (iv) The extra
accuracy yielded by a non-linear regression adjustment, over and above a
linear adjustment, is negligible.\footnote{The non-linear adjustment procedure
was also applied to the scores of the individual GARCH auxiliary models.
Again, the RMSE results were very similar to the corresponding results yielded
by the linear regression adjustment method and were thus not presented.} (v)
The score-based technique applied to simple auxiliary models from the GARCH
class, as long as allied with dimension reduction, yields reasonably accurate
estimates of the exact marginals and, in all but one case, more accurate
estimates than the ABC method in which summarization of any kind is avoided
(PMMH-ABC). }

\section{{Empirical illustration: Conditionally $\alpha$-stable returns with
stochastic volatility\label{empiric}}}

{{We complete the paper with a small empirical illustration, in which we
highlight }the {particular }benefit of using auxiliary likelihood-based ABC
{in the case where} the measurement density is not available in closed form
{and a PMMH method in which evaluation of this density ordinate is used (such
as the PMMH-BPF algorithm illustrated in Section \ref{sq_numerics}) is not
feasible.} {For the purpose of this illustration we} employ the stochastic
volatility model with conditionally $\alpha$-stable returns, defined in
(\ref{ret_2}) and (\ref{ar1}), using daily data on the S\&P500 index. {The
returns data (sourced from Reuters) extend} from 2 January 2013 to 7 February
2017, comprising 1033 observations, {and }are computed from open-to-close
prices. We standardize the returns by dividing each {observation} by {the}
{sample standard deviation, and fix} $\phi_{1}=0$. }

{We estimate the parameters $\phi_{2}$, $\phi_{3}$ and $\phi_{4}$ by auxiliary
{score}-based ABC, with the auxiliary model defined by the GARCH model in
(\ref{miss1_2}) and (\ref{miss2_2}), {and without the use of dimension
reduction techniques}. We elect to retain 500 ABC draws to estimate the ABC
posterior densities from 331,882 ABC replications (with quantile selection
guided by Frazier \textit{et al.}, 2018, as described earlier). Given the lack
of }a closed-form measurement density, {due to the $\alpha$-stable error term
in (\ref{ret_2}), the PMMH-ABCF algorithm, as described in Section \ref{data},
is applied as a comparator.\footnote{Note that it is possible to approximate
the $\alpha$-{stable measurement} density numerically {and, hence, exploit
this approximation in an application of a} PMMH-BPF algorithm, as described in
Section \ref{data}. However, {this} numerical {approximation is very
computationally burdensome:} a single likelihood evaluation for this model
using the \texttt{stabledist} package in \texttt{R} taking approximately half
an hour. {The unbiasedness of the estimated likelihood function under this
approximation would also have to be established, if one wished to claim that
the PMMH-BPF chain retained the correct invariant distribution.}} The three
{unknown} parameters are drawn as a block, using truncated normal proposals
formed based on the previous draw. The {marginal} posteriors are approximated
from 10,000 draws following 5,000 burn-in draws, {and with 3000 particles used
in each instance of likelihood estimation. }For both {ABC }methods, we employ
uniform priors: $\phi_{2}\sim U(0.7,0.999)$, $\phi_{3}\sim U(0.001,0.5)$, and
$\phi_{4}\sim U(1.2,2)$. {The MH acceptance ratio for the PMMH-ABCF for this
empirical illustration is 51\%.} Computation times for the {auxiliary
score-based} and {PMMH-ABCF} {algorithms} are 4 hours and 12.5 hours,
respectively. }

{Panels A, B and C of Figure \ref{post} depict the posterior {density
estimates for} $\phi_{2}$, $\phi_{3}$ and $\phi_{4}$, respectively, with the
{auxiliary score-based} {estimates} plotted {using a} solid line and those
{produced by} {PMMH-ABCF} plotted {with} {a dashed line}. The {PMMH-ABCF
estimate of the} posterior of $\phi_{2}$ is flat and mimics the prior, while
the corresponding {auxiliary score-based estimate} is well concentrated and
has a mode around {an empirically plausible value of }0.95. There is {also }a
stark contrast {between} the two posterior {estimates for} $\phi_{3}$, with
the score-based estimate peaking around 0.25, while the {PMMH-ABCF estimate}
peaks at the lower bound, suggesting a lack of convergence. The {two
estimated} posteriors for $\phi_{4}$, the parameter of the $\alpha$-{stable}
distribution, are {more similar one to the other}, both peaking at empirically
plausible values between 1.96-1.98. }

{{These }results {are encouraging evidence for the use of the auxiliary score
technique in an empirically relevant example in which the only other feasible
}}comparator{{ is both three times slower and}} produces results that do{{ not
appear to be {uniformly }informative or reliable.} }

{\small 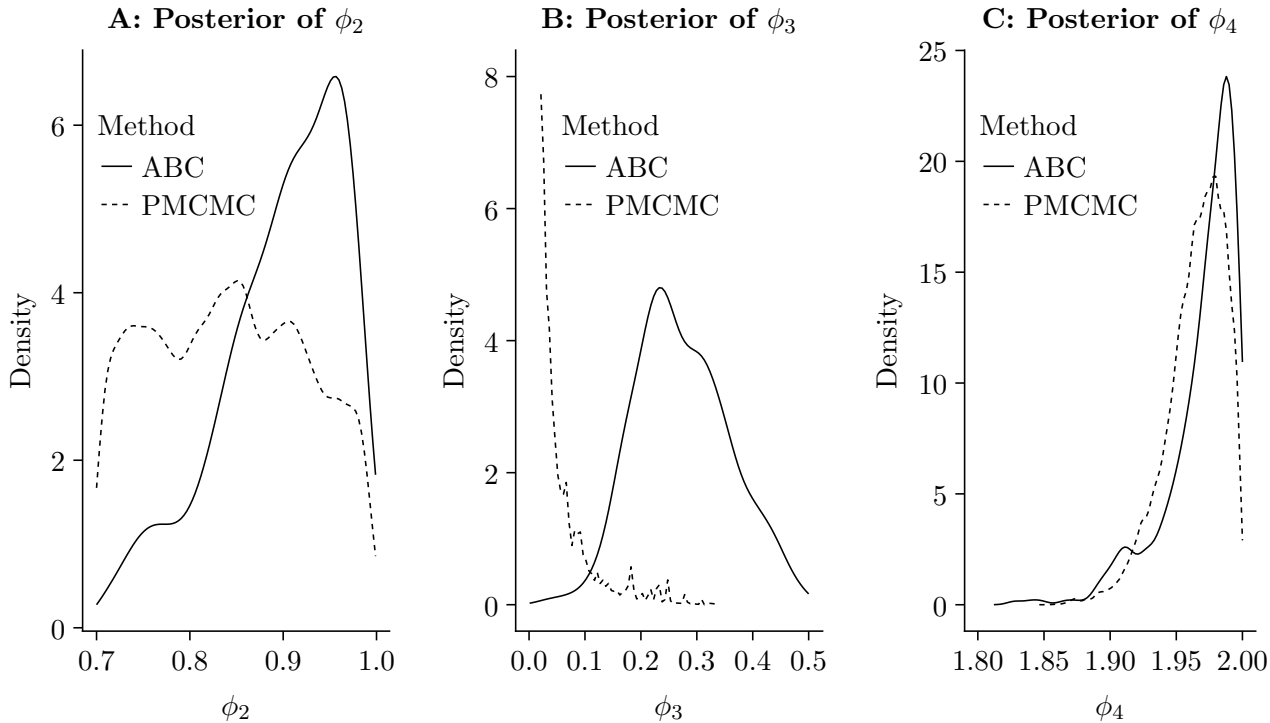
\begin{figure}[h]
{\small \centering
\input{postploth.tex}  }\caption{Posterior densities{ of {the }parameters} {of
the stochastic volatility model with conditionally }$\alpha${-stable returns
(defined in (\ref{ret_2}) and (\ref{ar1})) for the S\&P500 returns: }$\phi
_{2}$ (Panel A), $\phi_{3}$ (Panel B) and $\phi_{4}$ (Panel C). The posterior
densities generated by {the auxiliary score-based method} are plotted in solid
lines, while those generated by {PMMH-ABCF} are plotted in {dashed} lines.}%
\label{post}%
\end{figure}}

\section{{Conclusions and discussion\label{end}}}

{This paper has explored the application of approximate Bayesian computation
({ABC}) in the state space setting, in which auxiliary likelihood functions
are used to generate the matching statistics. Bayesian consistency of the
auxiliary likelihood-based method {- including for the computationally
efficient version based on the auxiliary score - }has been established, under
regularity conditions that exploit the state space structure of the auxiliary
model. The idea of tackling the dimensionality issue via an integrated
likelihood approach {has been} proposed and shown to yield {improved accuracy}
in our numerical experiments. {In a comprehensive numerical comparison with
alternative methods of estimating exact posterior densities, the
auxiliary-score based method is shown to perform very well, including in
comparison with particle filtering-based algorithms. Its ability to produce
plausible posterior estimates in an empirical setting is also demonstrated,}
with an alternative approach that applies ABC principles at the filtering
level yielding results that are not empirically sensible. }

{Despite the focus of this paper being on inference about the static
parameters in the state space model, there is nothing to preclude marginal
inference on the states being conducted, at a second stage. Specifically,
conditional on the (accepted) draws used to estimate $p(\mathbf{\phi|y})$,
existing filtering and smoothing methods (including{ methods\ that exploit}
ABC at the filtering level) could be used to yield draws of the states, and
(marginal) smoothed posteriors for the states produced via the usual averaging
arguments. {Such exploration is left for future work.} }

\bigskip

\noindent{\underline{\textbf{SUPPLEMENTARY MATERIAL:}} } The Supplementary
Appendix contains: 1) {the proofs of Theorems 1 and 2}; 2) an informal
demonstration of the equivalence of the auxiliary MLE-based approach to ABC
and the corresponding approach based on the auxiliary score; 3) additional
numerical exercises demonstrating the large sample behavior of ABC; and 4)
implementation details for the AUKF in the SV-SQ example. Gauss, C, Matlab
{and \texttt{{R}} }code used to produce all numerical results in the paper
{have also been provided on-line.}

\end{document}

%% file: postploth.tex
% Created by tikzDevice version 0.11 on 2018-10-15 09:15:47
% !TEX encoding = UTF-8 Unicode
\begin{tikzpicture}[x=1pt,y=1pt]
\definecolor{fillColor}{RGB}{255,255,255}
\path[use as bounding box,fill=fillColor,fill opacity=0.00] (0,0) rectangle (491.44,289.08);
\begin{scope}
\path[clip] ( 40.72, 42.90) rectangle (156.81,262.84);
\definecolor{drawColor}{RGB}{0,0,0}

\path[draw=drawColor,line width= 0.6pt,line join=round] ( 45.99, 52.90) --
	( 47.06, 54.40) --
	( 48.13, 55.98) --
	( 49.19, 57.63) --
	( 50.26, 59.33) --
	( 51.32, 61.07) --
	( 52.39, 62.85) --
	( 53.46, 64.68) --
	( 54.52, 66.52) --
	( 55.59, 68.39) --
	( 56.65, 70.25) --
	( 57.72, 72.09) --
	( 58.79, 73.86) --
	( 59.85, 75.56) --
	( 60.92, 77.13) --
	( 61.98, 78.56) --
	( 63.05, 79.81) --
	( 64.12, 80.88) --
	( 65.18, 81.75) --
	( 66.25, 82.41) --
	( 67.31, 82.88) --
	( 68.38, 83.18) --
	( 69.45, 83.33) --
	( 70.51, 83.38) --
	( 71.58, 83.37) --
	( 72.65, 83.37) --
	( 73.71, 83.43) --
	( 74.78, 83.62) --
	( 75.84, 84.00) --
	( 76.91, 84.62) --
	( 77.98, 85.52) --
	( 79.04, 86.74) --
	( 80.11, 88.30) --
	( 81.17, 90.23) --
	( 82.24, 92.52) --
	( 83.31, 95.17) --
	( 84.37, 98.17) --
	( 85.44,101.50) --
	( 86.50,105.14) --
	( 87.57,109.04) --
	( 88.64,113.17) --
	( 89.70,117.51) --
	( 90.77,122.00) --
	( 91.83,126.60) --
	( 92.90,131.27) --
	( 93.97,135.95) --
	( 95.03,140.60) --
	( 96.10,145.15) --
	( 97.16,149.56) --
	( 98.23,153.76) --
	( 99.30,157.72) --
	(100.36,161.42) --
	(101.43,164.88) --
	(102.50,168.10) --
	(103.56,171.14) --
	(104.63,174.07) --
	(105.69,176.96) --
	(106.76,179.89) --
	(107.83,182.91) --
	(108.89,186.09) --
	(109.96,189.45) --
	(111.02,192.98) --
	(112.09,196.66) --
	(113.16,200.42) --
	(114.22,204.19) --
	(115.29,207.87) --
	(116.35,211.35) --
	(117.42,214.55) --
	(118.49,217.42) --
	(119.55,219.90) --
	(120.62,222.03) --
	(121.68,223.84) --
	(122.75,225.43) --
	(123.82,226.93) --
	(124.88,228.46) --
	(125.95,230.16) --
	(127.02,232.12) --
	(128.08,234.41) --
	(129.15,237.04) --
	(130.21,239.97) --
	(131.28,243.06) --
	(132.35,246.15) --
	(133.41,248.97) --
	(134.48,251.24) --
	(135.54,252.64) --
	(136.61,252.84) --
	(137.68,251.54) --
	(138.74,248.47) --
	(139.81,243.44) --
	(140.87,236.36) --
	(141.94,227.24) --
	(143.01,216.21) --
	(144.07,203.51) --
	(145.14,189.49) --
	(146.20,174.56) --
	(147.27,159.19) --
	(148.34,143.84) --
	(149.40,128.96) --
	(150.47,114.94) --
	(151.53,102.08);

\path[draw=drawColor,line width= 0.6pt,dash pattern=on 2pt off 2pt ,line join=round] ( 45.99, 97.13) --
	( 47.06,111.36) --
	( 48.13,124.09) --
	( 49.19,134.25) --
	( 50.26,141.51) --
	( 51.32,146.24) --
	( 52.39,149.28) --
	( 53.46,151.43) --
	( 54.52,153.25) --
	( 55.59,154.97) --
	( 56.65,156.50) --
	( 57.72,157.65) --
	( 58.79,158.30) --
	( 59.85,158.50) --
	( 60.92,158.43) --
	( 61.98,158.28) --
	( 63.05,158.18) --
	( 64.12,158.14) --
	( 65.18,158.04) --
	( 66.25,157.76) --
	( 67.31,157.21) --
	( 68.38,156.37) --
	( 69.45,155.25) --
	( 70.51,153.90) --
	( 71.58,152.37) --
	( 72.65,150.73) --
	( 73.71,149.06) --
	( 74.78,147.53) --
	( 75.84,146.34) --
	( 76.91,145.74) --
	( 77.98,145.92) --
	( 79.04,146.94) --
	( 80.11,148.65) --
	( 81.17,150.79) --
	( 82.24,153.02) --
	( 83.31,155.08) --
	( 84.37,156.82) --
	( 85.44,158.27) --
	( 86.50,159.62) --
	( 87.57,161.11) --
	( 88.64,162.89) --
	( 89.70,164.97) --
	( 90.77,167.11) --
	( 91.83,169.02) --
	( 92.90,170.47) --
	( 93.97,171.48) --
	( 95.03,172.27) --
	( 96.10,173.10) --
	( 97.16,174.07) --
	( 98.23,175.02) --
	( 99.30,175.55) --
	(100.36,175.19) --
	(101.43,173.59) --
	(102.50,170.73) --
	(103.56,166.93) --
	(104.63,162.78) --
	(105.69,158.92) --
	(106.76,155.88) --
	(107.83,153.90) --
	(108.89,153.04) --
	(109.96,153.15) --
	(111.02,153.92) --
	(112.09,155.01) --
	(113.16,156.14) --
	(114.22,157.17) --
	(115.29,158.12) --
	(116.35,159.03) --
	(117.42,159.80) --
	(118.49,160.22) --
	(119.55,160.03) --
	(120.62,159.10) --
	(121.68,157.45) --
	(122.75,155.22) --
	(123.82,152.61) --
	(124.88,149.78) --
	(125.95,146.86) --
	(127.02,143.96) --
	(128.08,141.13) --
	(129.15,138.41) --
	(130.21,135.92) --
	(131.28,133.82) --
	(132.35,132.32) --
	(133.41,131.49) --
	(134.48,131.22) --
	(135.54,131.21) --
	(136.61,131.12) --
	(137.68,130.72) --
	(138.74,130.04) --
	(139.81,129.27) --
	(140.87,128.64) --
	(141.94,128.15) --
	(143.01,127.50) --
	(144.07,126.06) --
	(145.14,123.11) --
	(146.20,118.09) --
	(147.27,110.91) --
	(148.34,101.86) --
	(149.40, 91.64) --
	(150.47, 81.15) --
	(151.53, 71.27);
\end{scope}
\begin{scope}
\path[clip] (  0.00,  0.00) rectangle (491.44,289.08);
\definecolor{drawColor}{RGB}{0,0,0}

\path[draw=drawColor,line width= 0.6pt,line join=round,line cap=rect] ( 40.72, 42.90) --
	( 40.72,262.84);
\end{scope}
\begin{scope}
\path[clip] (  0.00,  0.00) rectangle (491.44,289.08);
\definecolor{drawColor}{RGB}{0,0,0}

\node[text=drawColor,anchor=base east,inner sep=0pt, outer sep=0pt, scale= 1] at ( 34.22, 40.04) {0};

\node[text=drawColor,anchor=base east,inner sep=0pt, outer sep=0pt, scale= 1] at ( 34.22,103.46) {2};

\node[text=drawColor,anchor=base east,inner sep=0pt, outer sep=0pt, scale=  1] at ( 34.22,166.87) {4};

\node[text=drawColor,anchor=base east,inner sep=0pt, outer sep=0pt, scale= 1] at ( 34.22,230.28) {6};
\end{scope}
\begin{scope}
\path[clip] (  0.00,  0.00) rectangle (491.44,289.08);
\definecolor{drawColor}{RGB}{0,0,0}

\path[draw=drawColor,line width= 0.6pt,line join=round] ( 37.22, 44.17) --
	( 40.72, 44.17);

\path[draw=drawColor,line width= 0.6pt,line join=round] ( 37.22,107.59) --
	( 40.72,107.59);

\path[draw=drawColor,line width= 0.6pt,line join=round] ( 37.22,171.00) --
	( 40.72,171.00);

\path[draw=drawColor,line width= 0.6pt,line join=round] ( 37.22,234.42) --
	( 40.72,234.42);
\end{scope}
\begin{scope}
\path[clip] (  0.00,  0.00) rectangle (491.44,289.08);
\definecolor{drawColor}{RGB}{0,0,0}

\path[draw=drawColor,line width= 0.6pt,line join=round,line cap=rect] ( 40.72, 42.90) --
	(156.81, 42.90);
\end{scope}
\begin{scope}
\path[clip] (  0.00,  0.00) rectangle (491.44,289.08);
\definecolor{drawColor}{RGB}{0,0,0}

\path[draw=drawColor,line width= 0.6pt,line join=round] ( 45.99, 39.40) --
	( 45.99, 42.90);

\path[draw=drawColor,line width= 0.6pt,line join=round] ( 81.29, 39.40) --
	( 81.29, 42.90);

\path[draw=drawColor,line width= 0.6pt,line join=round] (116.59, 39.40) --
	(116.59, 42.90);

\path[draw=drawColor,line width= 0.6pt,line join=round] (151.89, 39.40) --
	(151.89, 42.90);
\end{scope}
\begin{scope}
\path[clip] (  0.00,  0.00) rectangle (491.44,289.08);
\definecolor{drawColor}{RGB}{0,0,0}

\node[text=drawColor,anchor=base,inner sep=0pt, outer sep=0pt, scale=  1] at ( 45.99, 28.13) {0.7};

\node[text=drawColor,anchor=base,inner sep=0pt, outer sep=0pt, scale=  1] at ( 81.29, 28.13) {0.8};

\node[text=drawColor,anchor=base,inner sep=0pt, outer sep=0pt, scale=  1] at (116.59, 28.13) {0.9};

\node[text=drawColor,anchor=base,inner sep=0pt, outer sep=0pt, scale=  1] at (151.89, 28.13) {1.0};
\end{scope}
\begin{scope}
\path[clip] (  0.00,  0.00) rectangle (491.44,289.08);
\definecolor{drawColor}{RGB}{0,0,0}

\node[text=drawColor,anchor=base,inner sep=0pt, outer sep=0pt, scale=  1] at ( 98.76, 11.25) {$\phi_2$};
\end{scope}
\begin{scope}
\path[clip] (  0.00,  0.00) rectangle (491.44,289.08);
\definecolor{drawColor}{RGB}{0,0,0}

\node[text=drawColor,rotate= 90.00,anchor=base,inner sep=0pt, outer sep=0pt, scale=  1] at ( 20.93,152.87) {Density};
\end{scope}
\begin{scope}
\path[clip] (  0.00,  0.00) rectangle (491.44,289.08);

\path[] ( 40.72,191.49) rectangle (114.21,246.21);
\end{scope}
\begin{scope}
\path[clip] (  0.00,  0.00) rectangle (491.44,289.08);
\definecolor{drawColor}{RGB}{0,0,0}

\node[text=drawColor,anchor=base west,inner sep=0pt, outer sep=0pt, scale=  1] at ( 46.41,230.65) {Method};
\end{scope}
\begin{scope}
\path[clip] (  0.00,  0.00) rectangle (491.44,289.08);
\definecolor{drawColor}{RGB}{0,0,0}

\path[draw=drawColor,line width= 0.6pt,line join=round] ( 47.85,218.87) -- ( 59.42,218.87);
\end{scope}
\begin{scope}
\path[clip] (  0.00,  0.00) rectangle (491.44,289.08);
\definecolor{drawColor}{RGB}{0,0,0}

\path[draw=drawColor,line width= 0.6pt,dash pattern=on 2pt off 2pt ,line join=round] ( 47.85,204.41) -- ( 59.42,204.41);
\end{scope}
\begin{scope}
\path[clip] (  0.00,  0.00) rectangle (491.44,289.08);
\definecolor{drawColor}{RGB}{0,0,0}

\node[text=drawColor,anchor=base west,inner sep=0pt, outer sep=0pt, scale= 1] at ( 63.03,214.73) {ABC};
\end{scope}
\begin{scope}
\path[clip] (  0.00,  0.00) rectangle (491.44,289.08);
\definecolor{drawColor}{RGB}{0,0,0}

\node[text=drawColor,anchor=base west,inner sep=0pt, outer sep=0pt, scale=  1] at ( 63.03,200.28) {PMCMC};
\end{scope}
\begin{scope}
\path[clip] (  0.00,  0.00) rectangle (491.44,289.08);
\definecolor{drawColor}{RGB}{0,0,0}

\node[text=drawColor,anchor=base,inner sep=0pt, outer sep=0pt, scale=  1] at ( 98.76,271.13) {\bfseries A: Posterior of $\phi_2$};
\end{scope}
\begin{scope}
\path[clip] (204.53, 42.90) rectangle (320.62,262.84);
\definecolor{drawColor}{RGB}{0,0,0}

\path[draw=drawColor,line width= 0.6pt,line join=round] (209.81, 53.47) --
	(210.87, 53.68) --
	(211.94, 53.92) --
	(213.00, 54.18) --
	(214.07, 54.45) --
	(215.14, 54.73) --
	(216.20, 54.99) --
	(217.27, 55.24) --
	(218.33, 55.47) --
	(219.40, 55.70) --
	(220.47, 55.92) --
	(221.53, 56.16) --
	(222.60, 56.43) --
	(223.66, 56.74) --
	(224.73, 57.13) --
	(225.80, 57.60) --
	(226.86, 58.20) --
	(227.93, 58.96) --
	(228.99, 59.91) --
	(230.06, 61.09) --
	(231.13, 62.54) --
	(232.19, 64.28) --
	(233.26, 66.37) --
	(234.33, 68.85) --
	(235.39, 71.77) --
	(236.46, 75.19) --
	(237.52, 79.13) --
	(238.59, 83.59) --
	(239.66, 88.53) --
	(240.72, 93.85) --
	(241.79, 99.43) --
	(242.85,105.12) --
	(243.92,110.77) --
	(244.99,116.27) --
	(246.05,121.55) --
	(247.12,126.65) --
	(248.18,131.62) --
	(249.25,136.57) --
	(250.32,141.60) --
	(251.38,146.74) --
	(252.45,151.95) --
	(253.51,157.09) --
	(254.58,161.93) --
	(255.65,166.17) --
	(256.71,169.55) --
	(257.78,171.85) --
	(258.84,172.91) --
	(259.91,172.73) --
	(260.98,171.42) --
	(262.04,169.17) --
	(263.11,166.27) --
	(264.18,163.05) --
	(265.24,159.82) --
	(266.31,156.86) --
	(267.37,154.38) --
	(268.44,152.47) --
	(269.51,151.10) --
	(270.57,150.17) --
	(271.64,149.49) --
	(272.70,148.86) --
	(273.77,148.09) --
	(274.84,146.99) --
	(275.90,145.49) --
	(276.97,143.52) --
	(278.03,141.11) --
	(279.10,138.30) --
	(280.17,135.16) --
	(281.23,131.75) --
	(282.30,128.13) --
	(283.36,124.34) --
	(284.43,120.45) --
	(285.50,116.52) --
	(286.56,112.66) --
	(287.63,108.96) --
	(288.69,105.52) --
	(289.76,102.41) --
	(290.83, 99.66) --
	(291.89, 97.26) --
	(292.96, 95.15) --
	(294.03, 93.27) --
	(295.09, 91.55) --
	(296.16, 89.91) --
	(297.22, 88.30) --
	(298.29, 86.67) --
	(299.36, 84.97) --
	(300.42, 83.18) --
	(301.49, 81.27) --
	(302.55, 79.22) --
	(303.62, 77.06) --
	(304.69, 74.81) --
	(305.75, 72.52) --
	(306.82, 70.25) --
	(307.88, 68.06) --
	(308.95, 66.00) --
	(310.02, 64.10) --
	(311.08, 62.36) --
	(312.15, 60.79) --
	(313.21, 59.38) --
	(314.28, 58.12) --
	(315.35, 57.00);

\path[draw=drawColor,line width= 0.6pt,dash pattern=on 2pt off 2pt ,line join=round] (214.07,246.14) --
	(215.14,220.40) --
	(216.20,171.70) --
	(217.27,157.22) --
	(218.33,130.73) --
	(219.40,115.81) --
	(220.47,101.99) --
	(221.53, 96.03) --
	(222.60, 94.73) --
	(223.66, 99.17) --
	(224.73, 83.83) --
	(225.80, 75.35) --
	(226.86, 81.00) --
	(227.93, 79.77) --
	(228.99, 80.44) --
	(230.06, 71.72) --
	(231.13, 69.61) --
	(232.19, 65.79) --
	(233.26, 64.64) --
	(234.33, 62.11) --
	(235.39, 65.15) --
	(236.46, 60.92) --
	(237.52, 62.21) --
	(238.59, 59.88) --
	(239.66, 60.97) --
	(240.72, 58.28) --
	(241.79, 57.96) --
	(242.85, 58.29) --
	(243.92, 56.56) --
	(244.99, 57.60) --
	(246.05, 58.23) --
	(247.12, 60.14) --
	(248.18, 67.24) --
	(249.25, 58.48) --
	(250.32, 55.09) --
	(251.38, 55.71) --
	(252.45, 57.15) --
	(253.51, 54.50) --
	(254.58, 56.37) --
	(255.65, 58.58) --
	(256.71, 54.84) --
	(257.78, 59.01) --
	(258.84, 60.40) --
	(259.91, 53.97) --
	(260.98, 54.82) --
	(262.04, 62.28) --
	(263.11, 54.39) --
	(264.18, 53.23) --
	(265.24, 53.52) --
	(266.31, 53.47) --
	(267.37, 53.42) --
	(268.44, 56.74) --
	(269.51, 53.86) --
	(270.57, 53.31) --
	(271.64, 53.78) --
	(272.70, 53.12) --
	(273.77, 53.06) --
	(274.84, 54.97) --
	(275.90, 52.90) --
	(276.97, 53.45) --
	(278.03, 53.42) --
	(279.10, 53.30) --
	(280.17, 53.20);
\end{scope}
\begin{scope}
\path[clip] (  0.00,  0.00) rectangle (491.44,289.08);
\definecolor{drawColor}{RGB}{0,0,0}

\path[draw=drawColor,line width= 0.6pt,line join=round,line cap=rect] (204.53, 42.90) --
	(204.53,262.84);
\end{scope}
\begin{scope}
\path[clip] (  0.00,  0.00) rectangle (491.44,289.08);
\definecolor{drawColor}{RGB}{0,0,0}

\node[text=drawColor,anchor=base east,inner sep=0pt, outer sep=0pt, scale=1] at (198.03, 48.76) {0};

\node[text=drawColor,anchor=base east,inner sep=0pt, outer sep=0pt, scale=1] at (198.03, 98.75) {2};

\node[text=drawColor,anchor=base east,inner sep=0pt, outer sep=0pt, scale=1] at (198.03,148.74) {4};

\node[text=drawColor,anchor=base east,inner sep=0pt, outer sep=0pt, scale=1] at (198.03,198.72) {6};

\node[text=drawColor,anchor=base east,inner sep=0pt, outer sep=0pt, scale=1] at (198.03,248.71) {8};
\end{scope}
\begin{scope}
\path[clip] (  0.00,  0.00) rectangle (491.44,289.08);
\definecolor{drawColor}{RGB}{0,0,0}

\path[draw=drawColor,line width= 0.6pt,line join=round] (201.03, 52.90) --
	(204.53, 52.90);

\path[draw=drawColor,line width= 0.6pt,line join=round] (201.03,102.88) --
	(204.53,102.88);

\path[draw=drawColor,line width= 0.6pt,line join=round] (201.03,152.87) --
	(204.53,152.87);

\path[draw=drawColor,line width= 0.6pt,line join=round] (201.03,202.86) --
	(204.53,202.86);

\path[draw=drawColor,line width= 0.6pt,line join=round] (201.03,252.84) --
	(204.53,252.84);
\end{scope}
\begin{scope}
\path[clip] (  0.00,  0.00) rectangle (491.44,289.08);
\definecolor{drawColor}{RGB}{0,0,0}

\path[draw=drawColor,line width= 0.6pt,line join=round,line cap=rect] (204.53, 42.90) --
	(320.62, 42.90);
\end{scope}
\begin{scope}
\path[clip] (  0.00,  0.00) rectangle (491.44,289.08);
\definecolor{drawColor}{RGB}{0,0,0}

\path[draw=drawColor,line width= 0.6pt,line join=round] (209.59, 39.40) --
	(209.59, 42.90);

\path[draw=drawColor,line width= 0.6pt,line join=round] (230.74, 39.40) --
	(230.74, 42.90);

\path[draw=drawColor,line width= 0.6pt,line join=round] (251.90, 39.40) --
	(251.90, 42.90);

\path[draw=drawColor,line width= 0.6pt,line join=round] (273.05, 39.40) --
	(273.05, 42.90);

\path[draw=drawColor,line width= 0.6pt,line join=round] (294.20, 39.40) --
	(294.20, 42.90);

\path[draw=drawColor,line width= 0.6pt,line join=round] (315.35, 39.40) --
	(315.35, 42.90);
\end{scope}
\begin{scope}
\path[clip] (  0.00,  0.00) rectangle (491.44,289.08);
\definecolor{drawColor}{RGB}{0,0,0}

\node[text=drawColor,anchor=base,inner sep=0pt, outer sep=0pt, scale=1] at (209.59, 28.13) {0.0};

\node[text=drawColor,anchor=base,inner sep=0pt, outer sep=0pt, scale=1] at (230.74, 28.13) {0.1};

\node[text=drawColor,anchor=base,inner sep=0pt, outer sep=0pt, scale=1] at (251.90, 28.13) {0.2};

\node[text=drawColor,anchor=base,inner sep=0pt, outer sep=0pt, scale=1] at (273.05, 28.13) {0.3};

\node[text=drawColor,anchor=base,inner sep=0pt, outer sep=0pt, scale=1] at (294.20, 28.13) {0.4};

\node[text=drawColor,anchor=base,inner sep=0pt, outer sep=0pt, scale=1] at (315.35, 28.13) {0.5};
\end{scope}
\begin{scope}
\path[clip] (  0.00,  0.00) rectangle (491.44,289.08);
\definecolor{drawColor}{RGB}{0,0,0}

\node[text=drawColor,anchor=base,inner sep=0pt, outer sep=0pt, scale=1] at (262.58, 11.25) {$\phi_3$};
\end{scope}
\begin{scope}
\path[clip] (  0.00,  0.00) rectangle (491.44,289.08);
\definecolor{drawColor}{RGB}{0,0,0}

\node[text=drawColor,rotate= 90.00,anchor=base,inner sep=0pt, outer sep=0pt, scale=1] at (184.74,152.87) {Density};
\end{scope}
\begin{scope}
\path[clip] (  0.00,  0.00) rectangle (491.44,289.08);

\path[] (216.14,191.49) rectangle (289.63,246.21);
\end{scope}
\begin{scope}
\path[clip] (  0.00,  0.00) rectangle (491.44,289.08);
\definecolor{drawColor}{RGB}{0,0,0}

\node[text=drawColor,anchor=base west,inner sep=0pt, outer sep=0pt, scale=1] at (221.83,230.65) {Method};
\end{scope}
\begin{scope}
\path[clip] (  0.00,  0.00) rectangle (491.44,289.08);
\definecolor{drawColor}{RGB}{0,0,0}

\path[draw=drawColor,line width= 0.6pt,line join=round] (223.27,218.87) -- (234.84,218.87);
\end{scope}
\begin{scope}
\path[clip] (  0.00,  0.00) rectangle (491.44,289.08);
\definecolor{drawColor}{RGB}{0,0,0}

\path[draw=drawColor,line width= 0.6pt,dash pattern=on 2pt off 2pt ,line join=round] (223.27,204.41) -- (234.84,204.41);
\end{scope}
\begin{scope}
\path[clip] (  0.00,  0.00) rectangle (491.44,289.08);
\definecolor{drawColor}{RGB}{0,0,0}

\node[text=drawColor,anchor=base west,inner sep=0pt, outer sep=0pt, scale=1] at (238.45,214.73) {ABC};
\end{scope}
\begin{scope}
\path[clip] (  0.00,  0.00) rectangle (491.44,289.08);
\definecolor{drawColor}{RGB}{0,0,0}

\node[text=drawColor,anchor=base west,inner sep=0pt, outer sep=0pt, scale=1] at (238.45,200.28) {PMCMC};
\end{scope}
\begin{scope}
\path[clip] (  0.00,  0.00) rectangle (491.44,289.08);
\definecolor{drawColor}{RGB}{0,0,0}

\node[text=drawColor,anchor=base,inner sep=0pt, outer sep=0pt, scale=1] at (262.58,271.13) {\bfseries B: Posterior of $\phi_3$};
\end{scope}
\begin{scope}
\path[clip] (374.34, 42.90) rectangle (484.44,262.84);
\definecolor{drawColor}{RGB}{0,0,0}

\path[draw=drawColor,line width= 0.6pt,line join=round] (385.41, 52.92) --
	(386.42, 52.96) --
	(387.43, 53.03) --
	(388.44, 53.17) --
	(389.45, 53.39) --
	(390.46, 53.65) --
	(391.48, 53.92) --
	(392.49, 54.14) --
	(393.50, 54.26) --
	(394.51, 54.30) --
	(395.52, 54.31) --
	(396.53, 54.34) --
	(397.54, 54.43) --
	(398.55, 54.56) --
	(399.56, 54.66) --
	(400.57, 54.72) --
	(401.59, 54.71) --
	(402.60, 54.61) --
	(403.61, 54.42) --
	(404.62, 54.16) --
	(405.63, 53.88) --
	(406.64, 53.66) --
	(407.65, 53.57) --
	(408.66, 53.63) --
	(409.67, 53.83) --
	(410.68, 54.09) --
	(411.69, 54.35) --
	(412.71, 54.55) --
	(413.72, 54.66) --
	(414.73, 54.69) --
	(415.74, 54.66) --
	(416.75, 54.60) --
	(417.76, 54.56) --
	(418.77, 54.64) --
	(419.78, 54.97) --
	(420.79, 55.63) --
	(421.80, 56.66) --
	(422.82, 57.99) --
	(423.83, 59.49) --
	(424.84, 61.02) --
	(425.85, 62.49) --
	(426.86, 63.90) --
	(427.87, 65.27) --
	(428.88, 66.69) --
	(429.89, 68.22) --
	(430.90, 69.87) --
	(431.91, 71.57) --
	(432.93, 73.11) --
	(433.94, 74.24) --
	(434.95, 74.73) --
	(435.96, 74.52) --
	(436.97, 73.76) --
	(437.98, 72.84) --
	(438.99, 72.17) --
	(440.00, 72.07) --
	(441.01, 72.53) --
	(442.02, 73.33) --
	(443.04, 74.20) --
	(444.05, 75.04) --
	(445.06, 76.01) --
	(446.07, 77.39) --
	(447.08, 79.34) --
	(448.09, 81.81) --
	(449.10, 84.66) --
	(450.11, 87.76) --
	(451.12, 91.10) --
	(452.13, 94.73) --
	(453.15, 98.68) --
	(454.16,102.94) --
	(455.17,107.52) --
	(456.18,112.45) --
	(457.19,117.81) --
	(458.20,123.61) --
	(459.21,129.84) --
	(460.22,136.56) --
	(461.23,143.93) --
	(462.24,152.17) --
	(463.26,161.37) --
	(464.27,171.32) --
	(465.28,181.54) --
	(466.29,191.54) --
	(467.30,201.19) --
	(468.31,210.84) --
	(469.32,221.04) --
	(470.33,231.79) --
	(471.34,242.04) --
	(472.35,249.77) --
	(473.37,252.84) --
	(474.38,249.79) --
	(475.39,240.00) --
	(476.40,223.60) --
	(477.41,201.14) --
	(478.42,174.01) --
	(479.43,144.78);

\path[draw=drawColor,line width= 0.6pt,dash pattern=on 2pt off 2pt ,line join=round] (402.60, 52.90) --
	(403.61, 52.90) --
	(404.62, 52.91) --
	(405.63, 52.93) --
	(406.64, 52.95) --
	(407.65, 52.99) --
	(408.66, 53.07) --
	(409.67, 53.13) --
	(410.68, 53.18) --
	(411.69, 53.37) --
	(412.71, 53.65) --
	(413.72, 54.00) --
	(414.73, 54.63) --
	(415.74, 55.14) --
	(416.75, 54.91) --
	(417.76, 54.39) --
	(418.77, 54.40) --
	(419.78, 54.84) --
	(420.79, 54.99) --
	(421.80, 54.96) --
	(422.82, 55.45) --
	(423.83, 56.55) --
	(424.84, 57.51) --
	(425.85, 57.76) --
	(426.86, 57.89) --
	(427.87, 58.31) --
	(428.88, 58.75) --
	(429.89, 59.36) --
	(430.90, 60.35) --
	(431.91, 61.37) --
	(432.93, 62.63) --
	(433.94, 64.58) --
	(434.95, 66.58) --
	(435.96, 68.51) --
	(436.97, 71.01) --
	(437.98, 73.86) --
	(438.99, 76.61) --
	(440.00, 79.98) --
	(441.01, 83.73) --
	(442.02, 85.56) --
	(443.04, 86.06) --
	(444.05, 88.65) --
	(445.06, 93.11) --
	(446.07, 97.02) --
	(447.08,100.20) --
	(448.09,103.45) --
	(449.10,108.19) --
	(450.11,114.51) --
	(451.12,120.80) --
	(452.13,127.34) --
	(453.15,135.62) --
	(454.16,145.30) --
	(455.17,154.89) --
	(456.18,163.39) --
	(457.19,168.54) --
	(458.20,171.66) --
	(459.21,177.71) --
	(460.22,187.97) --
	(461.23,196.75) --
	(462.24,198.83) --
	(463.26,198.97) --
	(464.27,202.56) --
	(465.28,207.42) --
	(466.29,209.63) --
	(467.30,210.80) --
	(468.31,214.85) --
	(469.32,215.10) --
	(470.33,207.04) --
	(471.34,202.07) --
	(472.35,201.90) --
	(473.37,194.43) --
	(474.38,178.96) --
	(475.39,166.41) --
	(476.40,156.83) --
	(477.41,138.59) --
	(478.42,108.07) --
	(479.43, 77.26);
\end{scope}
\begin{scope}
\path[clip] (  0.00,  0.00) rectangle (491.44,289.08);
\definecolor{drawColor}{RGB}{0,0,0}

\path[draw=drawColor,line width= 0.6pt,line join=round,line cap=rect] (374.34, 42.90) --
	(374.34,262.84);
\end{scope}
\begin{scope}
\path[clip] (  0.00,  0.00) rectangle (491.44,289.08);
\definecolor{drawColor}{RGB}{0,0,0}

\node[text=drawColor,anchor=base east,inner sep=0pt, outer sep=0pt, scale=1] at (367.84, 48.76) {0};

\node[text=drawColor,anchor=base east,inner sep=0pt, outer sep=0pt, scale=1] at (367.84, 90.72) {5};

\node[text=drawColor,anchor=base east,inner sep=0pt, outer sep=0pt, scale=1] at (367.84,132.69) {10};

\node[text=drawColor,anchor=base east,inner sep=0pt, outer sep=0pt, scale= 1] at (367.84,174.65) {15};

\node[text=drawColor,anchor=base east,inner sep=0pt, outer sep=0pt, scale=1] at (367.84,216.61) {20};

\node[text=drawColor,anchor=base east,inner sep=0pt, outer sep=0pt, scale=1] at (367.84,258.57) {25};
\end{scope}
\begin{scope}
\path[clip] (  0.00,  0.00) rectangle (491.44,289.08);
\definecolor{drawColor}{RGB}{0,0,0}

\path[draw=drawColor,line width= 0.6pt,line join=round] (370.84, 52.89) --
	(374.34, 52.89);

\path[draw=drawColor,line width= 0.6pt,line join=round] (370.84, 94.86) --
	(374.34, 94.86);

\path[draw=drawColor,line width= 0.6pt,line join=round] (370.84,136.82) --
	(374.34,136.82);

\path[draw=drawColor,line width= 0.6pt,line join=round] (370.84,178.78) --
	(374.34,178.78);

\path[draw=drawColor,line width= 0.6pt,line join=round] (370.84,220.74) --
	(374.34,220.74);

\path[draw=drawColor,line width= 0.6pt,line join=round] (370.84,262.71) --
	(374.34,262.71);
\end{scope}
\begin{scope}
\path[clip] (  0.00,  0.00) rectangle (491.44,289.08);
\definecolor{drawColor}{RGB}{0,0,0}

\path[draw=drawColor,line width= 0.6pt,line join=round,line cap=rect] (374.34, 42.90) --
	(484.44, 42.90);
\end{scope}
\begin{scope}
\path[clip] (  0.00,  0.00) rectangle (491.44,289.08);
\definecolor{drawColor}{RGB}{0,0,0}

\path[draw=drawColor,line width= 0.6pt,line join=round] (379.34, 39.40) --
	(379.34, 42.90);

\path[draw=drawColor,line width= 0.6pt,line join=round] (404.37, 39.40) --
	(404.37, 42.90);

\path[draw=drawColor,line width= 0.6pt,line join=round] (429.39, 39.40) --
	(429.39, 42.90);

\path[draw=drawColor,line width= 0.6pt,line join=round] (454.41, 39.40) --
	(454.41, 42.90);

\path[draw=drawColor,line width= 0.6pt,line join=round] (479.43, 39.40) --
	(479.43, 42.90);
\end{scope}
\begin{scope}
\path[clip] (  0.00,  0.00) rectangle (491.44,289.08);
\definecolor{drawColor}{RGB}{0,0,0}

\node[text=drawColor,anchor=base,inner sep=0pt, outer sep=0pt, scale=1] at (379.34, 28.13) {1.80};

\node[text=drawColor,anchor=base,inner sep=0pt, outer sep=0pt, scale= 1] at (404.37, 28.13) {1.85};

\node[text=drawColor,anchor=base,inner sep=0pt, outer sep=0pt, scale=1] at (429.39, 28.13) {1.90};

\node[text=drawColor,anchor=base,inner sep=0pt, outer sep=0pt, scale=1] at (454.41, 28.13) {1.95};

\node[text=drawColor,anchor=base,inner sep=0pt, outer sep=0pt, scale=1] at (479.43, 28.13) {2.00};
\end{scope}
\begin{scope}
\path[clip] (  0.00,  0.00) rectangle (491.44,289.08);
\definecolor{drawColor}{RGB}{0,0,0}

\node[text=drawColor,anchor=base,inner sep=0pt, outer sep=0pt, scale=1] at (429.39, 11.25) {$\phi_4$};
\end{scope}
\begin{scope}
\path[clip] (  0.00,  0.00) rectangle (491.44,289.08);
\definecolor{drawColor}{RGB}{0,0,0}

\node[text=drawColor,rotate= 90.00,anchor=base,inner sep=0pt, outer sep=0pt, scale= 1] at (348.55,152.87) {Density};
\end{scope}
\begin{scope}
\path[clip] (  0.00,  0.00) rectangle (491.44,289.08);

\path[] (374.34,191.49) rectangle (447.84,246.21);
\end{scope}
\begin{scope}
\path[clip] (  0.00,  0.00) rectangle (491.44,289.08);
\definecolor{drawColor}{RGB}{0,0,0}

\node[text=drawColor,anchor=base west,inner sep=0pt, outer sep=0pt, scale=1] at (380.03,230.65) {Method};
\end{scope}
\begin{scope}
\path[clip] (  0.00,  0.00) rectangle (491.44,289.08);
\definecolor{drawColor}{RGB}{0,0,0}

\path[draw=drawColor,line width= 0.6pt,line join=round] (381.47,218.87) -- (393.04,218.87);
\end{scope}
\begin{scope}
\path[clip] (  0.00,  0.00) rectangle (491.44,289.08);
\definecolor{drawColor}{RGB}{0,0,0}

\path[draw=drawColor,line width= 0.6pt,dash pattern=on 2pt off 2pt ,line join=round] (381.47,204.41) -- (393.04,204.41);
\end{scope}
\begin{scope}
\path[clip] (  0.00,  0.00) rectangle (491.44,289.08);
\definecolor{drawColor}{RGB}{0,0,0}

\node[text=drawColor,anchor=base west,inner sep=0pt, outer sep=0pt, scale= 1] at (396.65,214.73) {ABC};
\end{scope}
\begin{scope}
\path[clip] (  0.00,  0.00) rectangle (491.44,289.08);
\definecolor{drawColor}{RGB}{0,0,0}

\node[text=drawColor,anchor=base west,inner sep=0pt, outer sep=0pt, scale=1] at (396.65,200.28) {PMCMC};
\end{scope}
\begin{scope}
\path[clip] (  0.00,  0.00) rectangle (491.44,289.08);
\definecolor{drawColor}{RGB}{0,0,0}

\node[text=drawColor,anchor=base,inner sep=0pt, outer sep=0pt, scale=1] at (429.39,271.13) {\bfseries C: Posterior of $\phi_4$};
\end{scope}
\end{tikzpicture}